\documentclass[aps,superscriptaddress,twocolumn,longbibliography]{revtex4-1}
\usepackage[]{graphicx}
\usepackage{amsmath,amssymb,amsfonts} 
\usepackage{pslatex}
\usepackage[]{subfigure}
\usepackage{multirow}
\usepackage{bm}
\usepackage{booktabs}
\newcommand{\fref}[1]{Fig.~\ref{#1}}
\newcommand{\sfref}[2]{Fig.~\ref{#1}\hyperref[#1]{#2}}
\newcommand{\sref}[2]{\ref{#1}\hyperref[#1]{#2}}
\renewcommand{\eqref}[1]{Eq.~(\ref{#1})}
\newcommand{\eref}[1]{(\ref{#1})}
\newcommand{\degree}{\ensuremath{^\circ}}

\usepackage[]{xcolor}
\definecolor{tirkizna}{rgb}{0.600, 0.000, 0.688}

\usepackage[sort&compress]{natbib}
\usepackage[pdfauthor={Mirta Herak},pdftitle={Site-Specific Spin Reorientation in AFM State of SeCuO3},colorlinks=true,citecolor=blue,linkcolor=blue,urlcolor=blue,bookmarks=true,bookmarksopen,bookmarksdepth=2]{hyperref}
\begin{document}
\title{\texorpdfstring{Site-Specific Spin Reorientation in Antiferromagnetic State of Quantum System ${\mathbf{SeCuO}}_{3}$}{Site-Specific Spin Reorientation in Antiferromagnetic State of Quantum System SeCuO3}}
\author{Nikolina~Novosel}
\affiliation{Institute of Physics, Bijeni\v{c}ka c. 46, HR-10000 Zagreb, Croatia}
\author{William~Lafargue-Dit-Hauret}
\affiliation{Institut des Sciences Chimiques de Rennes UMR 6226, Universit\'{e} de Rennes 1, Campus de Beaulieu, 35042 Rennes, France}
\author{\v{Z}eljko~Rapljenovi\'{c}}
\affiliation{Institute of Physics, Bijeni\v{c}ka c. 46, HR-10000 Zagreb, Croatia}
\author{Martina~Dragi\v{c}evi\'{c}}
\affiliation{Institute of Physics, Bijeni\v{c}ka c. 46, HR-10000 Zagreb, Croatia}
\author{Helmuth~Berger}
\affiliation{Institut de Physique de la Mati\`{e}re Complexe, EPFL, CH-1015 Lausanne, Switzerland}
\author{Dominik~Cin\v{c}i\'{c}}
\affiliation{Department of Chemistry, Faculty of Science, University of Zagreb, Horvatovac 102A, HR-10000 Zagreb, Croatia}
\author{Xavier~Rocquefelte}
\email{xavier.rocquefelte@univ-rennes1.fr}
\affiliation{Institut des Sciences Chimiques de Rennes UMR 6226, Universit\'{e} de Rennes 1, Campus de Beaulieu, 35042 Rennes, France}
\author{Mirta~Herak}
\email{mirta@ifs.hr}
\affiliation{Institute of Physics, Bijeni\v{c}ka c. 46, HR-10000 Zagreb, Croatia}


\date{\today}

\begin{abstract}
We report on the magnetocrystalline anisotropy energy (MAE) and spin reorientation in the antiferromagnetic (AFM) state of the spin $S=1/2$ tetramer system SeCuO$_3$ observed by torque magnetometry measurements in magnetic fields $H<5$~T and simulated using density functional calculations. We employ a simple phenomenological model of spin reorientation in finite magnetic field to describe our experimental torque data. Our results strongly support a collinear magnetic structure in zero field with a possibility of only very weak canting. Torque measurements also indicate that, contrary to what is expected for an uniaxial antiferromagnet, only fraction of the spins exhibit a spin-flop transition in SeCuO$_3$, allowing us to conclude that the AFM state of this system is unconventional and contains two decoupled subsystems. Our results demonstrate that the AFM state in SeCuO$_3$ is composed of a subsystem of AFM dimers, forming singlets, immersed in a long-range ordered AFM state, both states coexisting at the atomic scale. Furthermore, we show using ab-initio approach that both subsystems contribute differently to the MAE, corroborating the existence of decoupled subnetworks in SeCuO$_3$. The present combination of torque magnetometry, phenomenological and density functional theory approach to magnetic anisotropy represents a unique and original way to study site-specific reorientation phenomena in quantum magnets.
\end{abstract}
%\pacs{75.25.-j, 75.30.Gw,75.50.Ee}
%\keywords{}
\maketitle
\section{Introduction\label{sec:intro}}
\begin{figure}[tb]
	\centering
		\includegraphics[width=0.7\columnwidth]{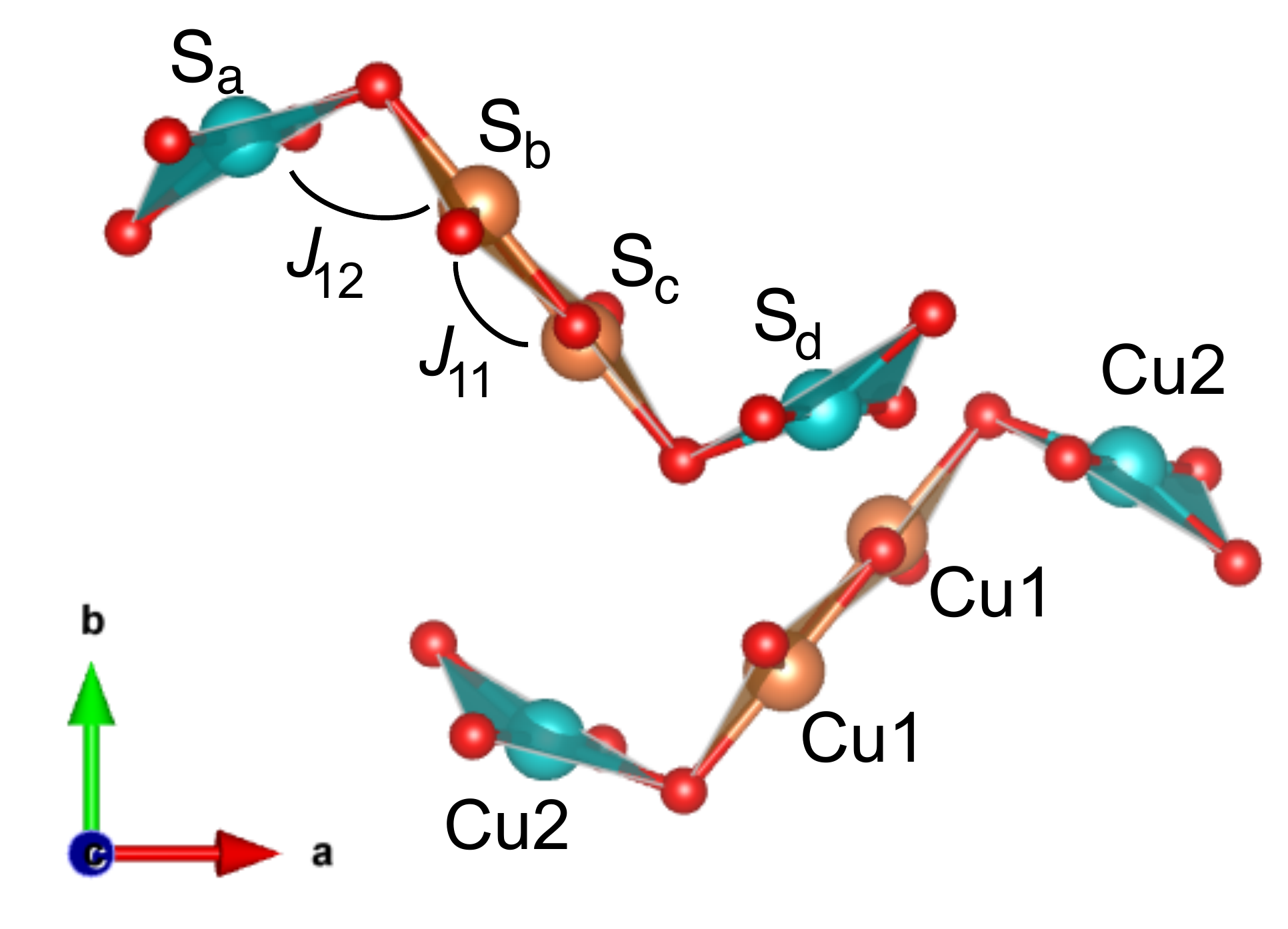}
	\caption{Two magnetically inequivalent tetramers in SeCuO$_3$ as proposed in Ref.~\onlinecite{Zivkovic-2012}. Cu1 atoms are shown in orange colour, Cu2 in blue and O atoms in red. Se atoms are not shown for simplicity.}
	\label{fig1}
\end{figure}

\indent Low-dimensional spin systems represent a fertile ground to study the influence of quantum effects on the formation of exotic states of matter. Zero-dimensional (0D) systems, in particular, are of significance since simple finite lattices represent a fruitful playground for theoretical investigations, while, at the same time, 0D magnetic lattices can be found in real materials allowing the theory to be tested. The simplest example of a 0D system is a spin dimer consisting of two spins coupled by exchange energy $J$. The two allowed states are singlet and triplet separated by an energy gap $J$ and the ground state is determined by the sign of $J$ (antiferromagnetic or ferromagnetic coupling). In real materials small, but finite, interactions between the 0D units can lead to a long range magnetic ordering. Combined with the quantum effects of underlying 0D magnetic units, this can lead to exotic phase diagrams where different phases can be obtained by tuning the relative strength of the exchange couplings.\\
\indent Another example of a 0D system is a spin tetramer where four spins $S_a$, $S_b$, $S_c$ and $S_d$ interact forming a 0D magnetic unit with slightly more complex excitation spectrum than found in  a spin dimer \cite{Haraldsen-2005}. When the coupling between spins $S_a$ and $S_b$ is equal to the coupling between spins $S_c$ and $S_d$, the spin Hamiltonian can be written as 
\begin{equation}\label{eq:Hamiltonian}
	\mathcal{H}= J_{12}\:(\mathbf{S}_a \cdot \mathbf{S}_b + \mathbf{S}_c \cdot \mathbf{S}_d) + J_{11}\:(\mathbf{S}_b \cdot \mathbf{S}_c).
\end{equation} 
An interesting limit for the spin tetramer is a case when the coupling $J_{11}$ between the two spins in the middle, $S_b$ and $S_c$, is antiferromagnetic (AFM)  and much stronger than the coupling $J_{12}$ of spins in the middle with spins on the sides of the tetramer, $S_a$ and $S_d$ (see \fref{fig1}). In this case $S_b$ and $S_c$ are expected to form a singlet state which persists in the background of  weakly connected paramagnetic spins $S_a$ and $S_d$. At low temperatures weak intertetramer interactions can lead to a long-range magnetic order. When this happens, the question arises whether singlet states are broken with $S_b$ and $S_c$ spins joining the long-range order, or do they somehow persist as singlets in the background of long-range magnetically ordered spins $S_a$ and $S_d$. The latter scenario was proposed by Hase \textit{et al.} for spin tetramer system CdCu$_2$(BO$_3$)$_2$ based on the high field magnetization measurements \cite{Hase-2005}, and it was recently confirmed by nuclear magnetic resonance (NMR) and zero-field muon spin relaxation (ZF-$\mu$SR) \cite{Lee-2014}. In the ordered AFM state of CdCu$_2$(BO$_3$)$_2$, spins S$_a$ and S$_d$ are related to Cu2 site, while S$_b$ and S$_c$ to Cu1 site. In this system, the spins of Cu1 atoms form strongly coupled singlets, but at T$_N$ = 9.8 K, antiferromagnetic long-range order sets in due to  much weaker intertetramer interactions. Neutron powder diffraction (NPD) measurements on this system reported significant magnetic moments on both Cu1 and Cu2 site, although it is smaller on Cu1 \cite{Hase-2009}. Detailed theoretical investigation of CdCu$_2$(BO$_3$)$_2$ showed dominant coupling between Cu1 spins forming AFM dimers, while further couplings forming intratetramer and intertetramer interactions are responsible for low-temperature AFM LRO of Cu2 spins which polarize Cu1 singlet states \cite{Janson-2012}. The same study revealed that polarization of Cu1 singlets is possible because the field from Cu2 spins is staggered and thus does not commute with the exchange interaction on the dimer \cite{Janson-2012}. Janson \textit{et al.} further showed that a significant magnetic moment can be induced on Cu1 spins by the staggered field from Cu2 spins \cite{Janson-2012}. This picture, later confirmed experimentally by NMR and ZF-$\mu$SR measurements \cite{Lee-2014}, is different from the usual long-range order which is induced by the interactions between the spins. In CdCu$_2$(BO$_3$)$_2$ magnetic interactions are responsible for the magnetic order of Cu2 site only, while Cu1 moments are polarized. The decoupling of the Cu1 and Cu2 spins in the ordered state is witnessed by a magnetic anomaly at $T^*=6.5$~K observed in NMR, which was attributed to the reorientation of Cu2 spins, while Cu1 remain intact \cite{Lee-2014}. To better understand how this type of ordering emerges it is important to find and study new systems which might host such exotic behavior. SeCuO$_3$ studied in this work is in many aspects similar to CdCu$_2$(BO$_3$)$_2$ and thus represents an ideal candidate to study site-selective quantum magnetism, especially since, unlike for the latter, high quality single crystals are available. \\
\indent  The monoclinic SeCuO$_3$ was recently proposed to be a host to a quantum linear spin tetramer system described by the Hamiltonian \eref{eq:Hamiltonian} where site-selective quantum correlations might play a significant role in establishing unusual magnetic properties \cite{Zivkovic-2012}.  SeCuO$_3$ crystals belong to the monoclinic space group $P2_{1}/n$ with unit cell parameters $a = 7.712$~\AA, $b = 8.238$~\AA, $c = 8.498$~\AA, and $\beta = 99.124^{\circ}$ \cite{Effenberger-1986}. Two crystallographically inequivalent copper sites, Cu1 and Cu2 are present in the monoclinic phase of SeCuO$_3$, each surrounded by six oxygen ligands forming a Jahn-Teller distorted elongated CuO$_6$ octahedra. This ligand configuration suggests $d_{x^2-y^2}$ orbital state for the unpaired copper spin $S=1/2$. Taking into account the local environment of the magnetic ion Cu$^{2+}$, it was proposed in Ref.~\onlinecite{Zivkovic-2012} that isolated linear spin tetramers Cu2-Cu1-Cu1-Cu2 are present in SeCuO$_3$. Two magnetically inequivalent spin tetramers in SeCuO$_3$ are shown in Fig. \ref{fig1}. However, the temperature dependence of the magnetic susceptibility cannot be explained by a simple tetramer Hamiltonian \eqref{eq:Hamiltonian} \cite{Emori-1975}, even if the $g$ factor temperature dependence is taken into account \cite{Herak-2014}. \\
\indent Previous studies of SeCuO$_3$ proposed that the rotation of macroscopic magnetic axes and the temperature variation of the electron $\mathbf{g}$ tensor in the paramagnetic state \cite{Zivkovic-2012,Herak-2014} are not a consequence of the temperature change of the crystal structure, but rather of a site-selective correlation which emerges from the large difference between  $J_{11}$ and $J_{12}$ couplings \cite{Zivkovic-2012}.  A recent nuclear quadrupole resonance (NQR) study proposed that Cu1 spins are strongly coupled forming a spin singlet state at temperatures $T<J_{11}\approx 200$~K \cite{Cvitanic-2018}.\\
\indent Below $T_N=8$~K, SeCuO$_3$ exhibits a long-range AFM order \cite{Zivkovic-2012,Herak-2014,Lee-2017,Cvitanic-2018}. Temperature dependence of the magnetic susceptibility anisotropy in low magnetic field, below $T_N$, is typical for uniaxial antiferromagnets \cite{Zivkovic-2012,Herak-2014}. A spin-flop transition is observed around $H_{SF}\approx 1.8$~T at $T=2$~K when the magnetic field is applied along the easy axis \cite{Zivkovic-2012}. Torque magnetometry reveals an unusual weak rotation of the magnetic axes with temperature in the $ac$ plane \cite{Zivkovic-2012,Herak-2014}, while NMR measurements demonstrate a possible rotation of the magnetic moments in the AFM state \cite{Lee-2017} connected to different moments developed on the Cu1 and Cu2 sites and different coupling between them. The high-field magnetization measurements showed the existence of a half-step magnetization plateau \cite{Lee-2017} similar to what was found in the previously mentioned CdCu$_2$(BO$_3$)$_2$ compound where the plateau emerges from the polarization of weakly coupled Cu2 spins while Cu1 dimers remain in the singlet state \cite{Hase-2005}. The $^{77}$Se NMR measurements revealed different temperature dependence of the Cu1 and Cu2 spin-spin relaxation rate $1/T_2$ in the AFM state which could be a signature of the presence of two subsystems in SeCuO$_3$, strongly coupled Cu1 dimers and weakly coupled Cu2 spins \cite{Lee-2017}. All these observations confirm that SeCuO$_3$ represents an ideal new host to study exotic magnetic behavior influenced by quantum phenomena.\\
\indent The magnetic structure of SeCuO$_3$ recently proposed from neutron powder diffraction measurements is highly noncollinear with different magnetic moment values at Cu1 and Cu2 sites, $m_{\textup{Cu1}}\approx 0.4\mu_B$ and $m_{\textup{Cu2}}\approx 0.7\mu_B$ at 1.5~K \cite{Cvitanic-2018}. The smaller value for Cu1 site, $m_{\textup{Cu1}}=0.35\mu_B$,  was confirmed by NQR measurements, but with significantly different orientation of magnetic moments with respect to  the CuO$_4$ plaquette than observed in the neutron diffraction experiment \cite{Cvitanic-2018}. Obviously, the AFM state in SeCuO$_3$ is more interesting than a simple uniaxial N\'{e}el state and calls for further studies. In this work we experimentally probe the magnetic anisotropy of the AFM state in SeCuO$_3$ using torque magnetometry measurements in magnetic fields $H\lesssim 5$~T significantly higher than the spin-flop field $H_{SF}\approx 1.8$~T. This allows us to determine the magnetocrystalline anisotropy energy (MAE) of SeCuO$_3$ as well as study the field-induced spin reorientation. We complete the description of the MAE by a theoretical investigation based on first-principles calculations.
%
%%%%%%%%%%%%%%%%%%%%%%%%%%%%%%%%%%%%%%%%%%%%%%%%%%%%%%%%%%%%%%%%%%%%%%%%%%%%%%%%%%%%%%%%%%%%%%%%%%%%%%%%%%%%%%%%%%
%
\section{Methods\label{sec:methods}}
\subsection{Experimental\label{sec:expmethods}}
\indent Single crystals of monoclinic SeCuO$_3$ have been grown by a standard chemical vapor phase method, as described in literature \cite{Zivkovic-2012}. \\
\indent The magnetic torque was measured by a home-built torque apparatus based on the torsion of a thin quartz fibre. The magnetic field was supplied by the Cryogenic Consultants 5~T split-coil superconducting magnet with a room-temperature bore. The quartz sample holder is placed in a separate cryostat which is mounted in the room-temperature bore of the magnet cryostat. The monitoring and control of the sample temperature were performed by the Lakeshore 336 temperature controller. For magnetic torque measurements a single crystal of mass $(246 \pm 8)~\mu$g was used with the $b$ axis parallel to the longest crystal axis and with the two crystal planes, $(1\:0\:1)$ and $(1\:0\:\overline{1})$, easily distinguishable. 
\subsection{Theory\label{sec:theorymethods}}
\indent The present calculations are based on the spin-polarized density functional theory as implemented in the Wien2k package \cite{blaha_w_2001} using a full potential linear augmented plane wave method. The Perdew-Burke-Ernzerhof approximation (PBE) \cite{perdew_generalized_1996} is considered for the exchange and correlation part. A Hubbard effective term within the Anisimov approach \cite{anisimov_density-functional_1993} was used to describe the Cu $3d$ orbitals more properly, allowing us to obtain magnetic moments for the copper sites close to 0.73 $\mu_B$. We also checked our calculations considering the on-site PBE0 hybrid functional \cite{tran_hybrid_2006}, giving similar observations. The R$_{MT}$ Muffin-Tin radii for Se, Cu and O atoms were set to 1.65, 1.96 and 1.49 bohr and the RK$_{max}$ to 6. The separation between valence and core states was set to -6 Ry, except for the calculations including zinc atoms (set to -8 Ry, with R$_{MT}$ = 1.96 bohr). The Brillouin zone sampling was done using a 5$\times$4$\times$4 k-mesh \cite{monkhorst_special_1976}.
\section{Results}\label{sec:results}
\subsection{Experimental Results}\label{sec:expresults}
\begin{figure}[tb]
	\centering
		\includegraphics[width=\columnwidth]{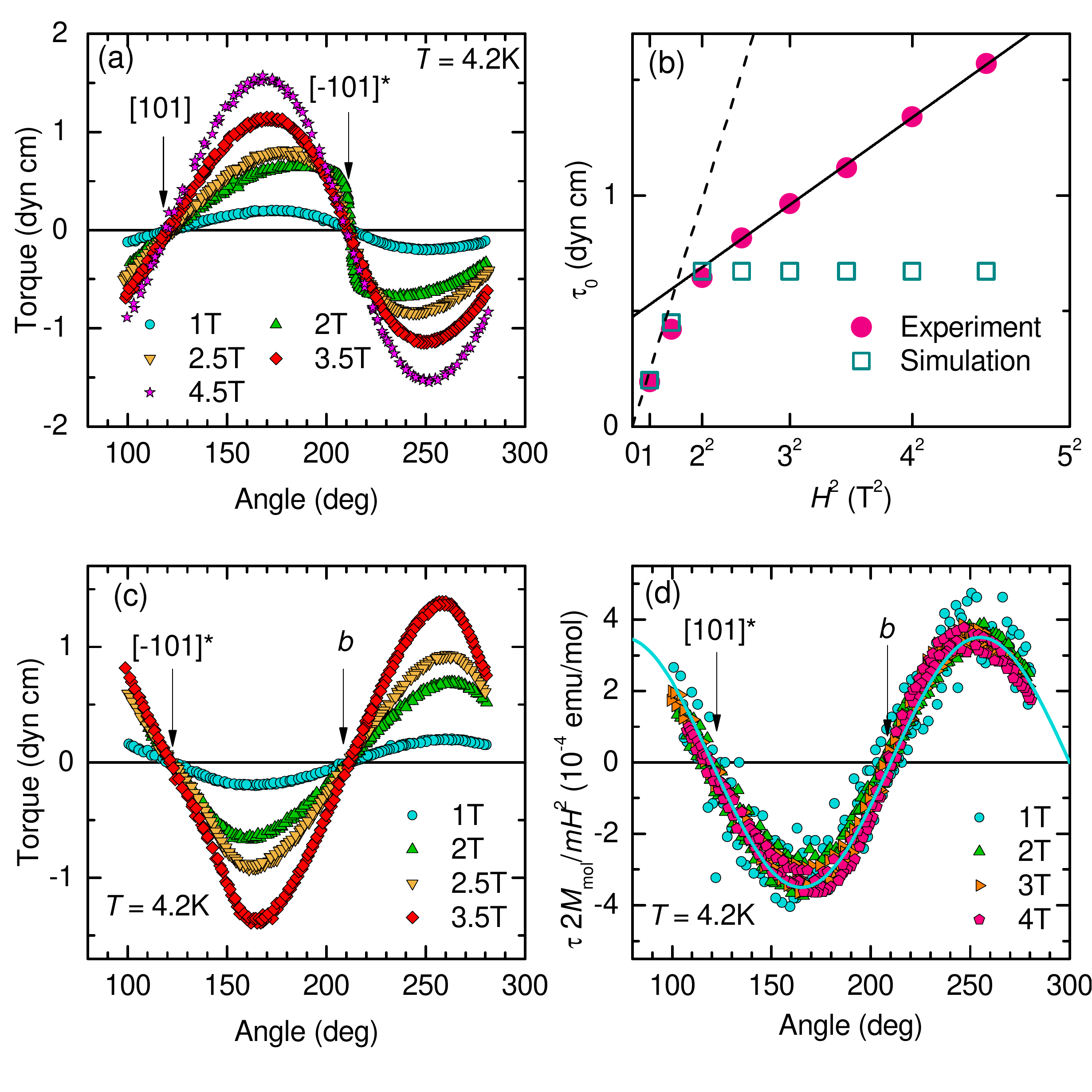}
	\caption{Angular dependence of torque $\tau$ measured in three crystal planes in different magnetic fields. (a) The torque measured in the $ac$ plane. (b) Full symbols: the dependence of torque amplitude $\tau_0$ on $H^2$ in the $ac$ plane [see (a)]. Solid line is fit to the experimental data for $H>H_{SF}$. Dashed line represents expected $H^2$ dependence of the torque amplitude for antiferromagnet with no reorientation (extrapolation of $H\ll H_{SF}$ data using \eqref{eq:torqueamplitude} and low-field anisotropy data \cite{Herak-2014}). Empty symbols represent simulation (see Sec. \ref{sec:simulation}). (c) Angular dependence of torque measured in the plane spanned by $b$ and $[\overline{1}\; 0\; 1]^*$ axes, and (d) the plane spanned by $b$ and  $[1\;0\;1]^*$ axes. In (d) the torque is multiplied by $2M_{mol}/ (mH^2)$ resulting in practically the same curve for all applied fields. This is consistent with no reorientation of spins in this plane [see \eqref{eq:measuredtorque}]. The angles corresponding to the specific crystal directions are pointed by arrows.}
	\label{fig2}
\end{figure}

\indent  For a simple collinear uniaxial antiferromagnet in low magnetic field $H\ll H_{SF}$, the magnetization is linearly dependent on the magnetic field. Consequently (see Appendix), angular dependence of the measured component of magnetic torque $\tau_z$ is then described by the expression
\begin{equation}\label{eq:measuredtorque}
\tau_z = \tau_0 \;\sin(2\varphi-2\varphi_0),
\end{equation}
where amplitude $\tau_0$ is given by
\begin{equation}\label{eq:torqueamplitude}
		\tau_0 = \dfrac{m}{2\;M_{mol}} \; \Delta \chi_{xy}\;H^2.
\end{equation}
$m$ is the mass of the sample, $M_{mol}$ is the molar mass and $H$ is the magnitude of applied magnetic field. $\Delta \chi_{xy} =\chi_{x}-\chi_{y}$ is the magnetic susceptibility anisotropy in the $xy$ plane in which the magnetic field rotates. $x$ and $y$ are, respectively, directions of the maximal and minimal susceptibility components in the plane of measurement. $\varphi$ is the goniometer angle and $\varphi_0$ is the angle at which the field is parallel to $x$. Eqs. \eref{eq:measuredtorque} and \eref{eq:torqueamplitude} show that, in the case of a linear response, the magnetic torque is proportional to $H^2$ and $\Delta \chi$ and the angular dependence of torque is a sine curve with a period of 180\degree. The previously published low-field ($H\lesssim 0.2$~T) torque data \cite{Herak-2014} in both paramagnetic and antiferromagnetic states are well described by Eqs. \eref{eq:measuredtorque} and \eref{eq:torqueamplitude}. \\
\indent The magnetic torque was measured in the AFM state at $T=4.2$~K by rotating the magnetic field in three crystal planes: the $ac$ plane, the plane spanned by $b$ and  $[\overline{1}\; 0\; 1]^*$ axes and the plane spanned by $b$ and $[1\; 0\; 1]^{*}$ axes. Angular dependence of torque for these three planes is shown in \fref{fig2}.  With the exception of the data shown in \sfref{fig2}{(d)} for plane spanned by  $b$ and $[1\; 0\; 1]^{*}$ axes, the measured torque cannot be described by \eqref{eq:measuredtorque}. The torque curves in Figs. \sref{fig2}{(a)} and \sref{fig2}{(c)} are not regular sine curves and the amplitude of torque does not increase linearly with $H^2$, as can be seen for the $ac$ plane in \sfref{fig2}{(b)}. The deviation of the torque curves from \eqref{eq:measuredtorque} is most pronounced in the $ac$ plane for $H=2$~T which is close to the spin-flop field $H_{SF}\approx 1.8$~T, see \sfref{fig2}{(a)}. \\
\indent For the plane spanned by $b$ and $[1\; 0\; 1]^{*}$ axes we show torque multiplied by $2M_{mol}/(m H^2)$ in \sfref{fig2}{(d)}. In this way the torque measured in different fields gives practically the same curve, which is expected from Eqs. \eref{eq:measuredtorque} and \eref{eq:torqueamplitude} and implies that there is no spin reorientation in this plane. In this work we show that the deviation from  low-field behaviour observed in the two other crystal planes is a result of the spin axis reorientation induced by magnetic field comparable in magnitude to the spin-flop field. \\
\indent In \sfref{fig2}{(b)} we plot the dependence of torque amplitude $\tau_0$ on $H^2$ measured in the $ac$ plane. The dashed line represents the slope given by $m/2\;M_{mol}\; \Delta \chi_{xy} $ [see \eqref{eq:torqueamplitude}] with $\Delta \chi$ obtained from the low-field measurements \cite{Herak-2014}. The deviation from the low-field behavior is observed already for $H\gtrsim 1.5$~T. Interestingly, the $H^2$ dependence seems to be restored for $H\geq 2$~T but with a much smaller slope which would, according to \eqref{eq:torqueamplitude}, correspond to a weaker susceptibility anisotropy $\Delta \chi$. This result is crucial for our interpretation of the spin reorientation in SeCuO$_3$ and in the following sections we will demonstrate how the observed behavior is a direct consequence of the site-specific spin reorientation induced by the magnetic field, comparable to or larger than the spin-flop field.
%
%%%%%%%%%%%%%%%%%%%%%%%%%%%%%%%%%%%%%%%%%%%%%%%%%%%%%%%%%%%%%%%%%%%%%%%%%%%%%%%%%%%%%%%%%%%%%%%%%%%%%%%%%%%%%%%%%%
%\clearpage
%
%
\subsection{\texorpdfstring{Phenomenological model of spin reorientation in ${\mathrm{SeCuO}}_{3}$}{Phenomenological model of spin reorientation in SeCuO3}}\label{sec:simulation}
\begin{figure}[tb]
	\centering
		\includegraphics[width=0.7\columnwidth]{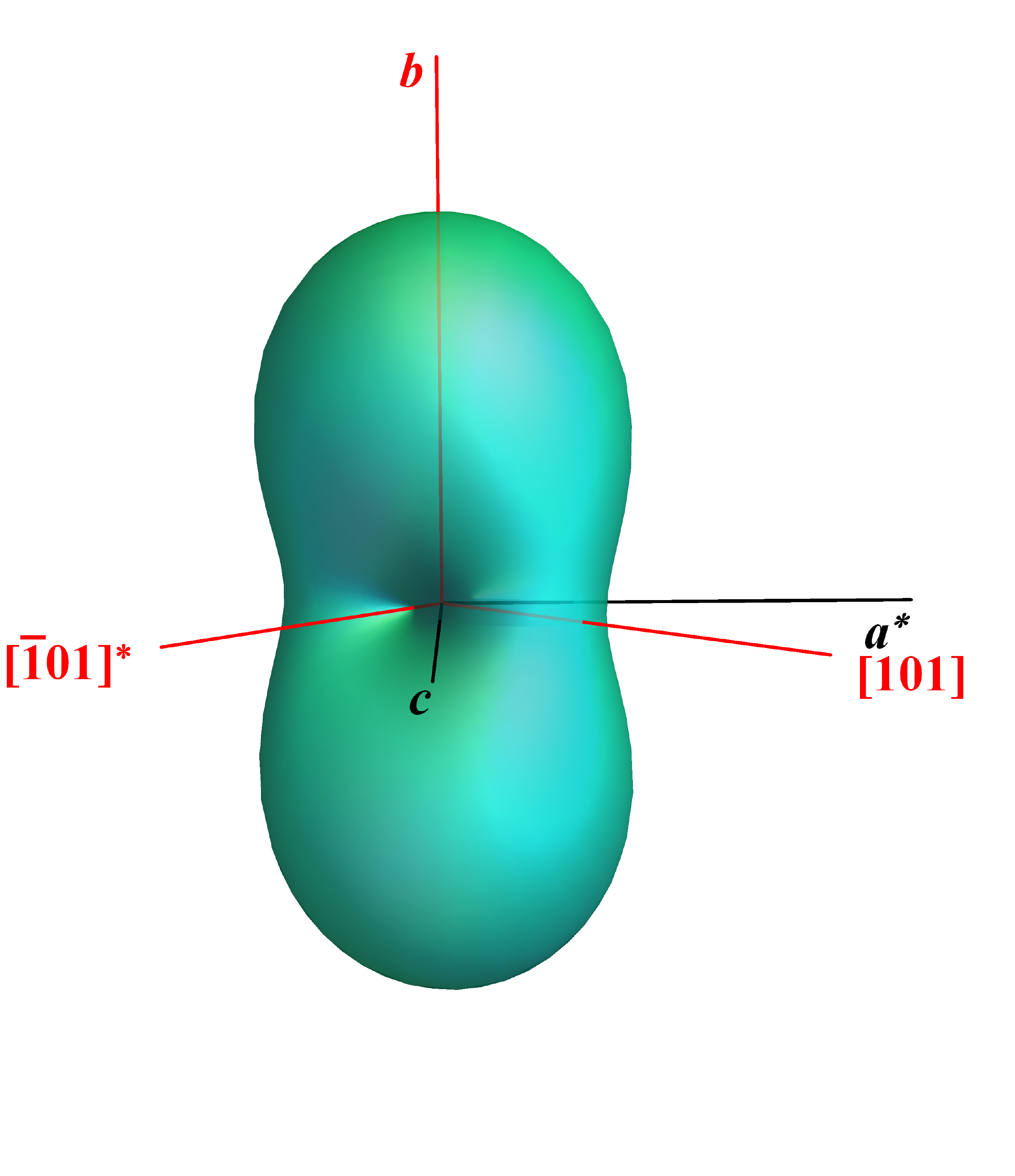}
	\caption{The MAE shape in SeCuO$_3$ obtained from the torque measurements in this work. Red lines represent the magnetic axes which are also extrema of the anisotropy energy. Easy axis direction (a global minimum) is along the $\left<\overline{1}\:0\:1\right>^*$ axis.}
	\label{fig3}
\end{figure}

\begin{figure}[tb]
	\centering
		\includegraphics[width=\columnwidth]{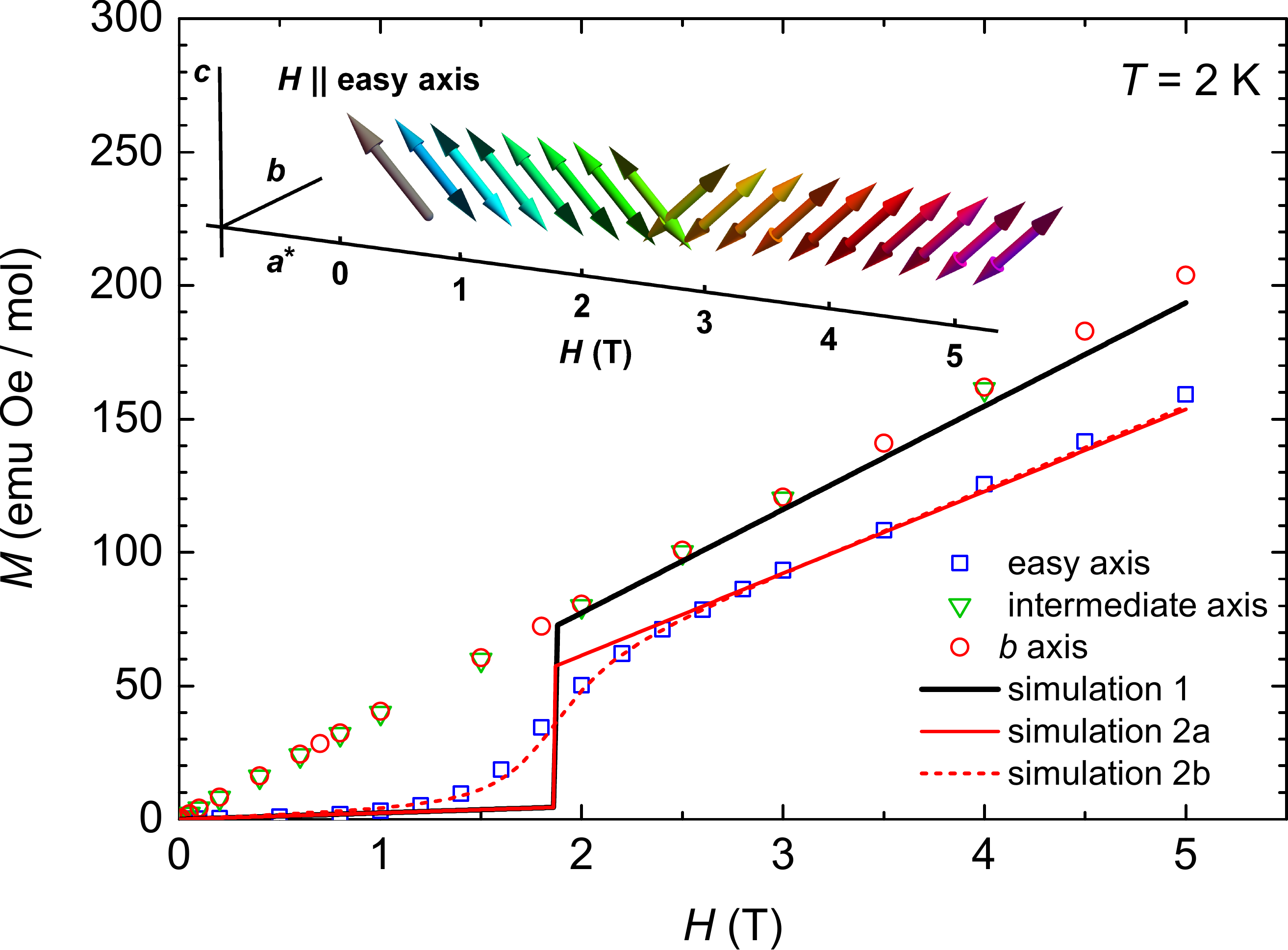}
	\caption{A simulation of the magnetization dependence on magnetic field when $H$ is parallel to the easy axis compared to the experimental result at $T=2$~K published in Ref.~\onlinecite{Zivkovic-2012} (empty blue squares). Inset: The dependence of the spin axis direction on applied magnetic field obtained from simulation. Simulation 1 allows all spins to rotate simultaneously. Simulation 2 allows only fraction of the spins to rotate, as described in the main text. Solid red line (2a) represents result for $H\parallel$ easy axis, while dashed line (2b) simulates a misorientation of the sample by 10\degree, to mimic probable misorientation in our experiment.}
	\label{fig4}
\end{figure}

\begin{figure}[tb]
	\centering
		\includegraphics[width=\columnwidth]{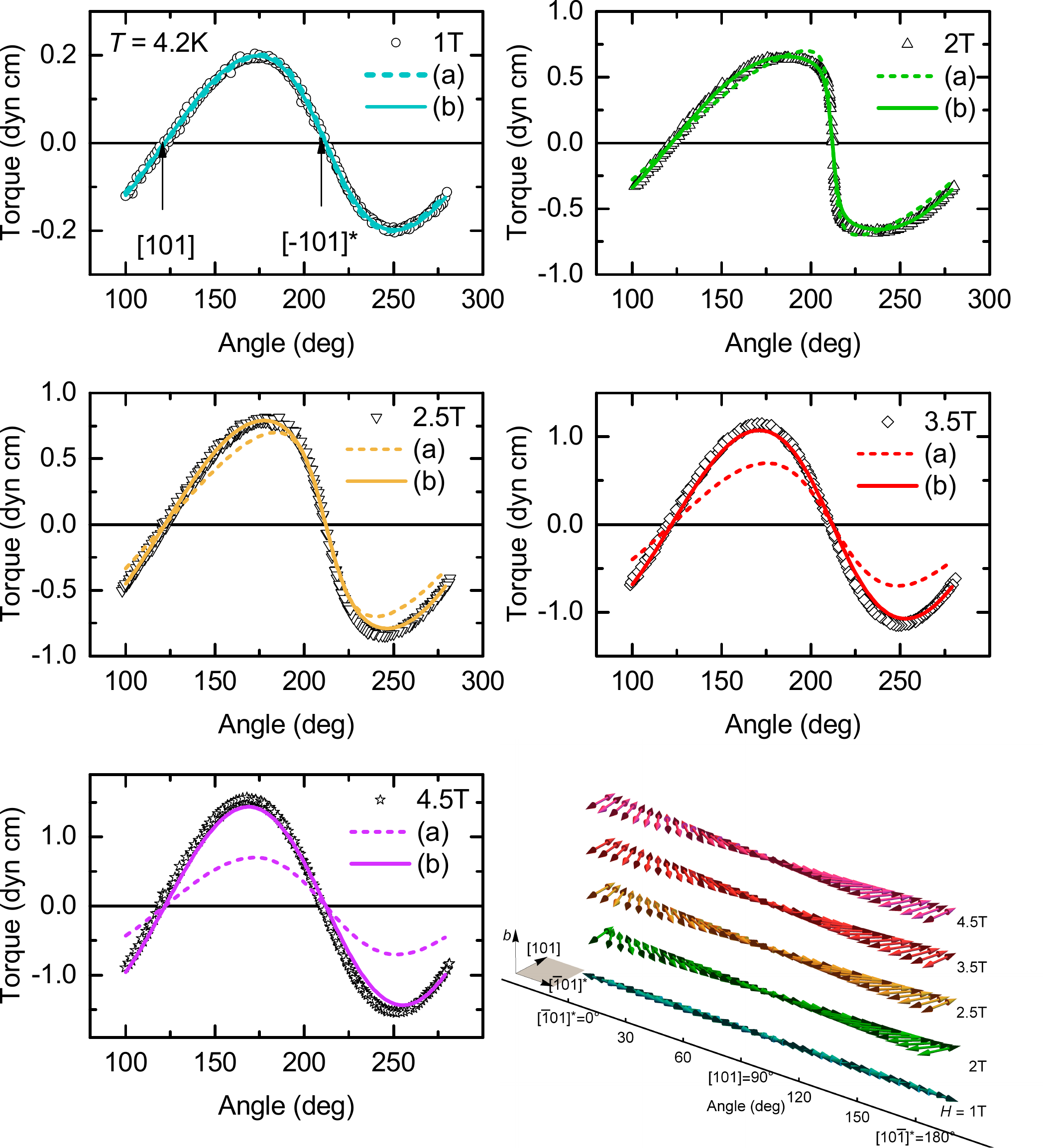}
	\caption{The torque measured in the $ac$ plane in different magnetic fields compared to two different simulations, as described in the main text. (a) A simulation with reorientation of all spins. (b) A simulation allowing the site-specific reorientation. Bottom right panel: the angle-dependent reorientation of the spin axis obtained from the simulation shown in laboratory coordinate system. A plane of rotation of the magnetic field is shown as dark square in the accompanying coordinate system. The angle is measured with respect to the $b$ axis, while in other panels a goniometer angle is shown. }
	\label{fig5}
\end{figure}

\begin{figure}[tb]
	\centering
		\includegraphics[width=\columnwidth]{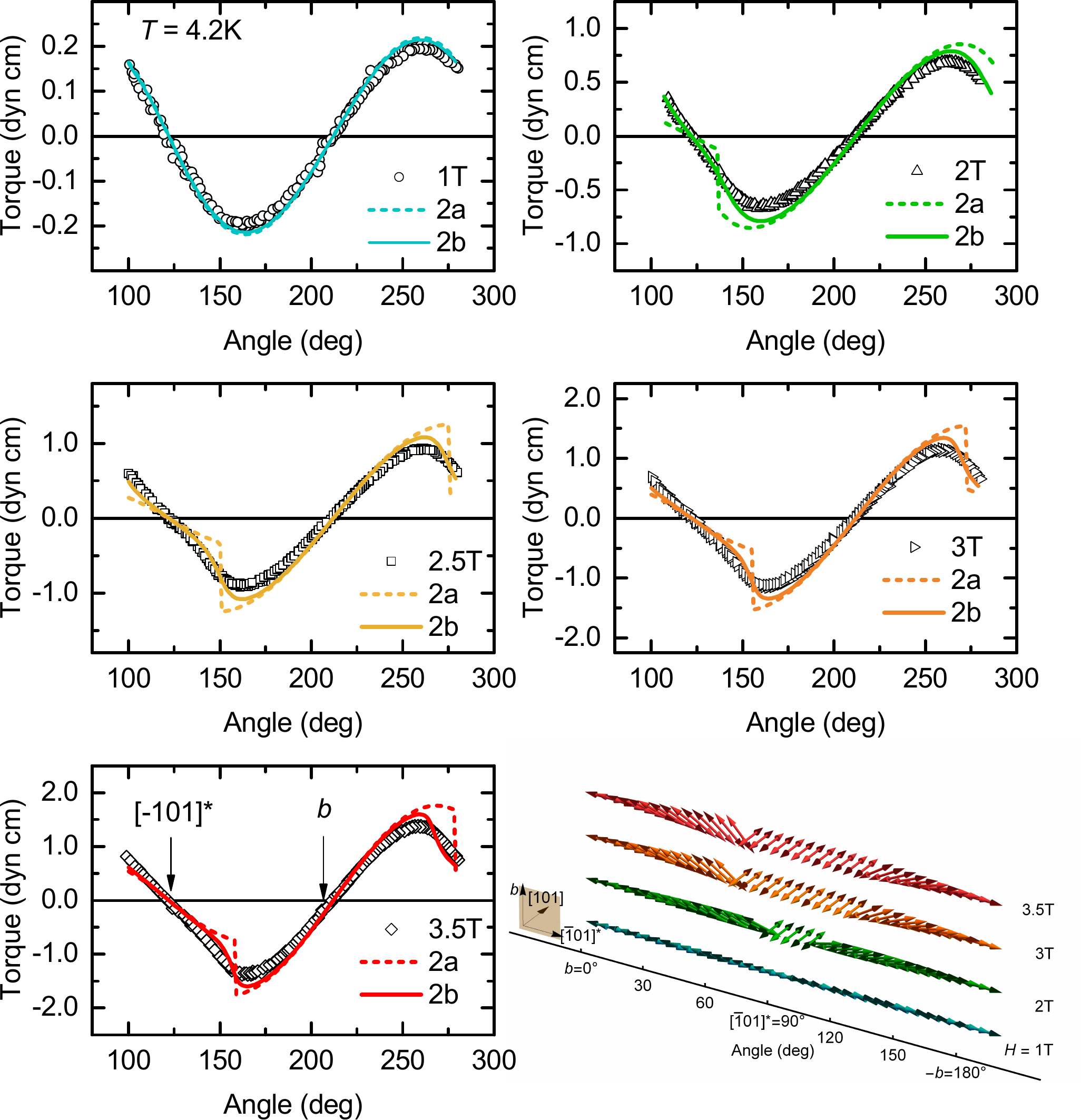}
	\caption{The torque measured in the plane spanned by $b$ and $[\overline{1}\:0\:1]^*$ axes compared to the results of simulation, as described in the main text. Dotted lines represent a perfect orientation (2a), and solid lines simulate a slightly misoriented sample (2b). Only site-specific simulation is shown. Bottom right panel: angle-dependent spin axis reorientation for perfect orientation of the sample shown in the laboratory coordinate system. The plane of rotation of the magnetic field is shown as a dark square in the accompanying coordinate system. The angle is measured with respect to the $b$ axis, while in other panels goniometer angle is shown. }
	\label{fig6}
\end{figure}

\begin{figure}[tb]
	\centering
		\includegraphics[width=\columnwidth]{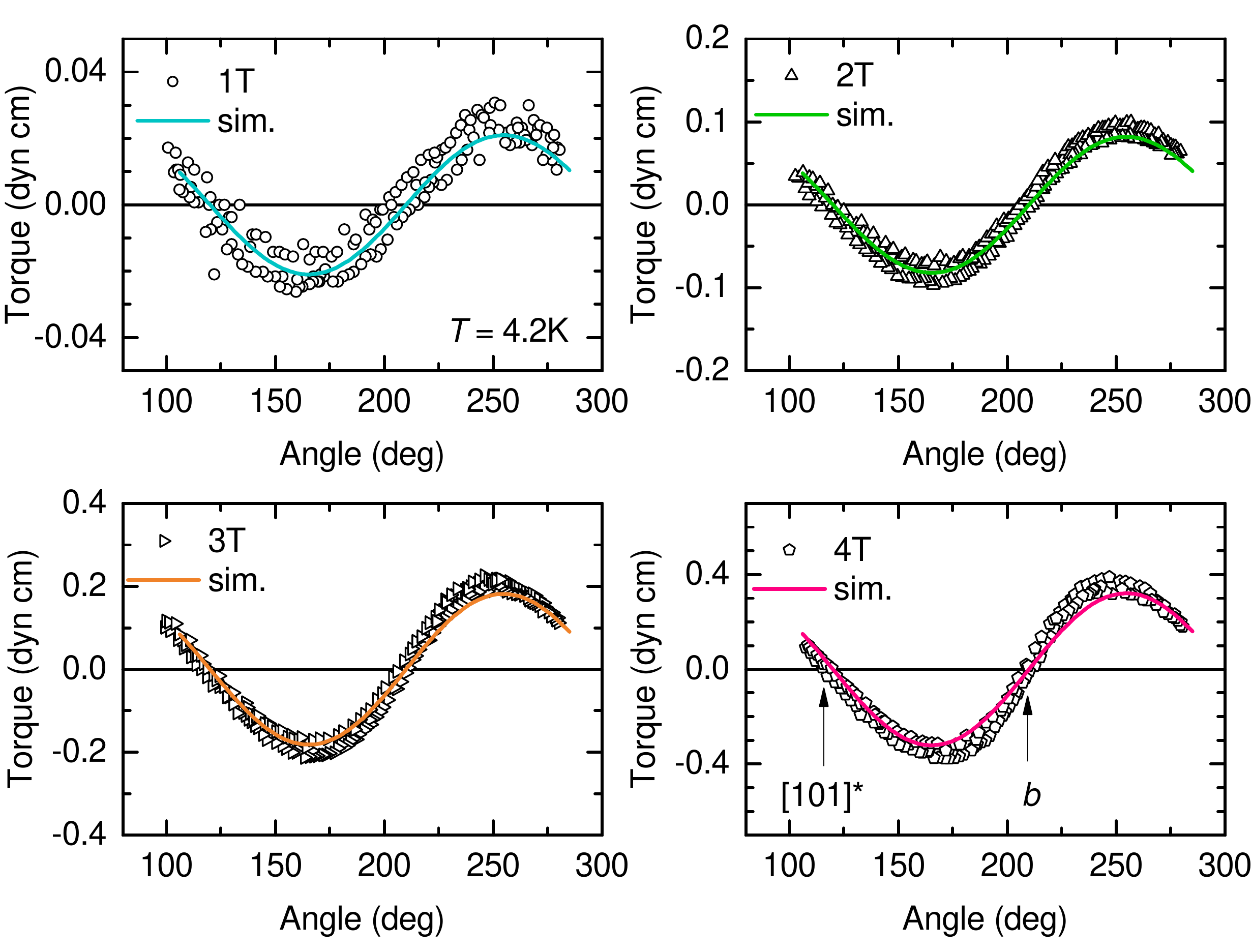}
	\caption{The torque measured in the plane spanned by $b$ and $[1\:0\:1]^*$ axes compared to  the results of the site-specific simulation (sim.), as described in the main text.}
	\label{fig7}
\end{figure}

\indent The spin reorientation in a collinear antiferromagnet in a finite applied magnetic field was first proposed by N\'{e}el in 1936 \cite{Neel-1936,Neel-1952} who observed that the competition between the orientation of spins defined by the magnetocrystalline anisotropy energy (MAE) and the orientation preferred by the Zeeman energy results in a reorientation of spins in such a way to minimize the total energy. The MAE determines the spin orientation in the absence of magnetic field (easy axis direction), while Zeeman energy for an antiferromagnet is minimal when spins are perpendicular to the applied magnetic field. Depending on the magnitude and direction of the applied field, spins will be oriented in such a way to minimize the total energy while still maintaining an almost collinear AFM structure due to the exchange energy which is much larger than the anisotropy energy. When magnetic field is applied along the easy axis, the spins reorient perpendicularly to magnetic field when the field reaches a critical value  $H_{SF}$ called the \emph{spin-flop} field. The magnitude of the spin-flop field depends on the magnitude of magnetocrystalline anisotropy energy, as well as the magnetic susceptibility anisotropy \cite{Rohrer-1969}. Spin-flop transitions were observed in many antiferromagnets and studied in detail in literature (see e.g. Ref.~\onlinecite{Bogdanov-2007} and references therein). A simple phenomenological approach proposed already by N\'{e}el can be used to study field-induced spin reorientation in antiferromagnets with different symmetries. Specifically, torque magnetometry measurements can be employed to determine the MAE shape and also to study the spin axis reorientation in a finite magnetic field. We have successfully used this approach previously to study the magnetocrystalline anisotropy in a uniaxial antiferromagnet \cite{Herak-2015}, but also in antiferromagnets with higher symmetries and multiple antiferromagnetic domains \cite{Herak-2010, Herak-2011}. \\
\indent To employ such a phenomonological approach to study the macroscopic spin reorientation in collinear antiferromagnets we need to know the magnetic susceptibility tensor, as well as the MAE. In accordance with Neumann's principle both the susceptibility tensor and MAE of SeCuO$_3$ must obey point group symmetry elements of $P2_{1}/n$ space group.\\
\indent The magnetic susceptibility tensor can be determined by torque measurements in low magnetic field, as shown in Appendix. The temperature dependence of the magnetic susceptibility anisotropy of SeCuO$_3$ in AFM state \cite{Zivkovic-2012, Herak-2014} is typical for an uniaxial antiferromagnet: susceptibility goes to zero as temperature decreases from $T_N$ to zero when magnetic field is applied along the easy axis, and is almost temperature-independent along the hard and the intermediate axes. Our previous torque measurements in low magnetic field \cite{Herak-2014} have shown that in the AFM state  below $T\approx 6$~K the easy magnetic axis is along $\left<\overline{1}\:0\:1 \right>^*$, while the intermediate and hard axes are along $\left<1\:0\:1 \right>$ and $\left<0\:1\:0 \right>=\pm b$ axes, respectively. This allows us to write the magnetic susceptibility tensor of SeCuO$_3$ in the AFM state at $T\lesssim 6$~K as
\begin{equation}\label{eq:tensorcsodia}
	\bm{\hat{\chi}}_0 = 
	\begin{bmatrix}
	\chi_{[\overline{1}\: 0 \:1]^*} & 0 & 0\\
	0 & \chi_{[1\:0\:1]} & 0 \\
	0 & 0 &  \chi_{b}
	\end{bmatrix}.
		\end{equation}
The coordinate system spanned by the magnetic eigenaxes in the AFM state below $\approx 6$~K is $([\overline{1}\: 0 \:1]^*,\; [1\:0\:1],\; b)$. This is in accordance with the symmetry requirements which dictate that the $b$ axis must be one of the magnetic eigenaxes, while the other two eigenaxes can have any orientation in the $ac$ plane \cite{Newnham}. One might note here that the measurement of susceptibility tensor allows the detection of symmetry breaking. A signature of symmetry lowering in monoclinic SeCuO$_3$ would be observed if the $b$ axis would no longer represent one of the magnetic eigenaxes, as explained in the Appendix. We have performed detailed measurements to determine if there is such a symmetry breaking in SeCuO$_3$ and, within the small error induced by a possible misorientation of the sample, our results show that symmetry is preserved in the AFM state of SeCuO$_3$, in agreement with the recent neutron powder diffraction and NQR measurements \cite{Cvitanic-2018}. Small rotation of the magnetic axes observed for 6~K$\lesssim T \lesssim T_N=8$~K \cite{Zivkovic-2012,Herak-2014} is confined to the $ac$ plane and is thus symmetry-preserving. \\
\indent Next, we write down the MAE, which also must satisfy the point group symmetry operations for SeCuO$_3$, as follows
\begin{equation}\label{eq:Fa}
	\mathcal{F}_a(\theta,\; \phi) = K_1 \; \cos^2 \theta + K_{2} \sin^2 \theta \cos 2\phi,
\end{equation}
where $\theta$ and $\phi$ are polar and azimuthal angles in the spherical coordinate system, and $K_1$ and $K_{2}$ are the anisotropy constants expressed in units of \emph{erg}/\emph{mol}. In principle, the anisotropy energy $\mathcal{F}_a$ can be written to higher-order terms and can also include extrinsic contributions to the anisotropy such as a shape anisotropy. However, second order terms, written in \eqref{eq:Fa}, are sufficient to study the reorientation of the spin axis in a finite magnetic field, as long as we do not try to describe critical behaviour in magnetic fields very close to the spin-flop field \cite{Bogdanov-2007}. \eqref{eq:Fa} describes the amount of energy needed to rotate the spin axis away from the easy axis direction. Depending on the value and sign of the anisotropy constants (both can be positive or negative), the anisotropy energy \eref{eq:Fa} can have several different shapes allowed by symmetry. Torque magnetometry is a technique which can be employed to experimentally determine the actual shape of the MAE for a specific crystal. As already mentioned, our previous low-field torque results along with the results shown in \fref{fig2} show that the easy axis direction in the AFM phase for $T\lesssim 6$~K  is along the $\left< \overline{1}\:0\:1\right>^*$ direction. This is then the direction of the global minimum of the anisotropy energy \eref{eq:Fa}. So, the other two extrema must be along the $\left<1\:0\:1\right>$ and $\pm b$ directions. The torque results presented in this work show that the global maximum i.e. hard axis is along the $\pm b$ direction, while the intermediate axis is  found along the $\left<1\:0\:1\right>$ direction. This puts the following constraints on the anisotropy constants: $K_2<0$ and $K_2 < K_1$. In \fref{fig3} we plot the resulting magnetocrystalline anisotropy energy in SeCuO$_3$. \\
In a finite magnetic field, $\mathbf{H}(\psi, \xi)$, the Zeeman energy is given by
\begin{equation}\label{eq:Zeeman}
	\mathcal{F}_{Z}(\psi, \xi) = -\dfrac{1}{2}\; \mathbf{H}(\psi, \xi) \cdot \bm{\hat{\chi}}\cdot \mathbf{H}(\psi, \xi)
\end{equation}
where $\bm{\hat{\chi}}$ is the magnetic susceptibility tensor of the sample expressed in the same coordinate system as the MAE \eref{eq:Fa}, and $\psi$ and $\xi$ are polar and azimuthal angles in the spherical coordinate system representing the direction of the magnetic field $\mathbf{H}= H\: (\cos\xi\sin\psi, \sin\xi\sin\psi, \cos\psi)$. In very low magnetic field the susceptibility tensor $\bm{\hat{\chi}}$ is given by \eref{eq:tensorcsodia}. In finite magnetic field, the spin axis will in general start to rotate away from the direction of the easy axis. We describe this rotation by allowing the susceptibility tensor to rotate
\begin{equation}\label{eq:Rotchi}
\bm{\hat{\chi}}(\theta,\phi) = \mathbf{R}(\theta,\phi)\cdot \bm{\hat{\chi}}_0 \cdot \mathbf{R}^T(\theta,\phi)
\end{equation}
where $\bm{\hat{\chi}}_0$ is the low-field ($H\ll H_{SF}$) susceptibility tensor given by the expression \eref{eq:tensorcsodia} and $\mathbf{R}(\theta,\phi)$ is the rotation matrix. Our torque measurements were performed at $T=4.2$~K where $\chi_{[\overline{1}\: 0 \:1]^*}=4\cdot10^{-4}$~emu/mol, $\chi_{[1\:0\:1]}=3.5\cdot10^{-3}$~emu/mol and $\chi_b= 3.8\cdot 10^{-3}$~emu/mol are the eigenvalues of the susceptibility tensor obtained from our previous susceptibility and torque measurements \cite{Zivkovic-2012,Herak-2014} .\\
\indent The total phenomenological energy $\mathcal{F}_{tot}$ of the sample in finite magnetic field is the sum of MAE and Zeeman energy
\begin{equation}\label{eq:Ftot}
	\mathcal{F}_{tot} (\theta, \phi, \psi, \xi)= \mathcal{F}_a (\theta, \phi) + \mathcal{F}_Z(\psi, \xi, \theta, \phi).
\end{equation}
For an anisotropic antiferromagnetic sample placed in finite magnetic field $\mathbf{H}(\psi, \xi)$, the new direction of the spin axis is obtained numerically by minimizing the total energy $\mathcal{F}_{tot}$ with respect to $\theta$ and $\phi$. To simulate the experimental results we start by finding $K_1$ and $K_2$ values which satisfy the above mentioned requirements for the MAE extrema. A correct choice of values must reproduce the experimental  $H_{SF}$ value of $\approx 1.8$~T at $T=2$~K when the magnetic field is applied along the easy axis. In the present case, the spin-flop field is given by $H_{SF}= [2\:(K_1-K_2)/(\chi_{[101]}-\chi_{[\overline{1}01]^*})]^{1/2}$ \cite{Uozaki-2000,Bogdanov-2007}. By simulating magnetic field dependence of magnetization to reproduce the experimental $H_{SF}$ value we can only pinpoint the difference $K_1-K_2$. The values of $K_1$ and $K_2$ will thus be obtained from the simulation of the angular dependence of torque data. \\
\indent We proceed by simulating the field dependence of magnetization for magnetic field applied along the easy axis direction at $T=2$~K and by comparing this result to the experimental data. Setting $K_2 - K_1 = -6.35\cdot 10^5$~erg/mol gives $H_{SF}=1.87$~T for values of the susceptibility tensor measured at $T=2$~K \cite{Herak-2014}. We fix our spin-flop field to this value since it is well within the margin of error for experimental data, and it also reproduces well our other results. We apply magnetic field ${\bf H}(\psi_0,\xi_0)$ along the easy axis by setting $\psi_0=\pi/2$, $\xi_0=0$ in the chosen coordinate system and then minimize numerically the total energy \eref{eq:Ftot} for each value of the applied field to obtain $\theta_0$ and $\phi_0$. Using the values obtained in this manner we calculate a rotated susceptibility tensor \eref{eq:Rotchi} and finally the magnetization from ${\bf M} =\bm{\hat{\chi}}(\theta_0,\phi_0) \cdot {\bf H}(\psi_0,\xi_0) $. In order to simulate the dc magnetization measurements we only take a component of the magnetization along the applied magnetic field. The result of our calculation, shown by solid black line in \fref{fig4} (simulation 1), is compared to the measured values from Ref.~\onlinecite{Zivkovic-2012} represented by empty blue squares in \fref{fig4}. The spin-flop transition is clearly observed at $H_{SF}=1.87$~T in the calculated curve. \\
\indent Our simulation also gives a new direction of the spin axis which for $H\geq H_{SF}$ rotate from the easy axis direction to the intermediate axis direction $\left<101\right>$, as shown in inset of \fref{fig4}. However, for $H>H_{SF}$, the calculated magnetization (solid black line in \fref{fig4}) is somewhat larger than the measured one (empty blue squares) and in fact falls on values obtained for magnetization measured along the intermediate and the hard axes (empty green triangles and red circles in \fref{fig4}). We should mention here that our result agrees with behaviour that is usually obtained in the experimental spin-flop. Significantly smaller magnetization observed in the experiment is in fact anomalous and suggests that the spin reorientation in SeCuO$_3$ might be unconventional. We will return to this point later. Experimental value of the spin-flop field is the same at 4.2~K as at 2~K, within the experimental uncertainty \cite{Zivkovic-2012}. Since susceptibility tensor components are slightly different at 4.2~K, we need to take $K_2-K_1=-5.42\cdot 10^5$~erg/mol in order to reproduce $H_{SF}\approx 1.87$~T at $T=4.2$~K. \\
\indent Finally, we proceed with calculating the angular dependence of torque for all planes probed during the measurements. We simulate a rotation of magnetic field by appropriately changing $\psi$ and $\xi$ for $\bf{H}(\psi,\xi)$. By minimizing the total energy \eref{eq:Ftot} we obtain the rotated susceptibility tensor from which we calculate magnetization $\bf{M}$ expressed in emu Oe/mol, as described above. Then we calculate torque from ${\bf \tau} =m/M_{mol}\; \bf{M} \times \bf{H}$, and plot only the component perpendicular to the plane of rotation of magnetic field, since only that component is measured in experiment.\\
\indent The results obtained for the $ac$ plane shown by dashed lines are compared to the measured torque curves in \fref{fig5}. For the $ac$ plane, our simulations agree very well with measurements for $H\leq 2$~T. However, at higher magnetic fields the calculated curves have smaller amplitude than the measured curves, and the discrepancy increases as the field increases. In order to understand the discrepancy between the calculated and the measured amplitudes $\tau_0$, we compare measurements  with the calculated $H^2$ dependence of $\tau_0$ in the $ac$ plane in \sfref{fig2}{(b)}. The measured amplitude increases nonlinearly with $H^2$ in low magnetic field, and for $H>2$~T the increase becomes linear with $H^2$, as we already pointed out in Section \ref{sec:expresults}. In comparison, the calculated torque amplitude follows the measured one below $H\lesssim 2$~T and then it becomes constant for $H\geq H_{SF}$ [empty squares in \sfref{fig2}{(b)}]. Independence of the torque amplitude on magnetic field for $H\geq H_{SF}$ obtained from our calculation is also observed in torque experiment for conventional spin reorientation in uniaxial collinear antiferromagnet \cite{Tokumoto-2005,Uozaki-2000} and is in agreement with the results of N\'{e}el \cite{Neel-1936, Neel-1952}.  Our experimental result points to a more exotic spin reorientation in SeCuO$_3$. Thus, in the remainder of this section we consider possible reasons for the deviation of the measured data from the results expected for a conventional uniaxial collinear antiferromagnet.\\
\indent The deviation of our experimental curves from the calculated ones could be a consequence of the lowering of the symmetry. This symmetry lowering would result in the MAE shape different form the one shown in \fref{fig3} and consequently could influence the measured angular dependencies of torque. However, the lowering of the symmetry can be disregarded, as we explained earlier.\\
\indent Another possibility for the observed discrepancy is a misorientation of the sample. A simulation of misorientation of the sample showed that the increase in torque amplitude mimicking the measured one can be reproduced only for an unrealistic assumption of misorientations of $\approx 50\degree$ or so, which were certainly not realized in experiment. \\
\indent Having discarded the symmetry lowering and the misorientation as possible reasons for discrepancy of the measured and the simulated data, we turn again to the torque amplitude dependence on $H^2$ shown with full circles in \sfref{fig2}{(b)}. According to \eqref{eq:torqueamplitude}, we expect the amplitude to increase linearly with $H^2$ when there is no spin reorientation, i.e. in an antiferromagnet for $H\ll H_{SF}$. If all the spins in the sample would remain oriented along the easy axis in applied field, the torque amplitude would increase with $H^2$ by following the dashed line in \sfref{fig2}{(b)} which represents the low-field behavior. Instead, for $H\geq H_{SF}$, the data can be fitted to $\tau_0 = a+b\cdot H^2$. The result of the fit, shown by solid line in \sfref{fig2}{(b)}, gives $a=(0.47 \pm 0.1)$~dyn~cm and $b=(0.0539 \pm 0.0009)$~dyn~cm /T$^2$. Using \eqref{eq:torqueamplitude},  from the fitting constant $b$ we obtain  $\Delta \chi = (8.4 \pm 0.1) \cdot 10^{-4}$~emu/mol. This value is significantly smaller than the total susceptibility anisotropy measured in the $ac$ plane at 4.2~K in low magnetic field, $\Delta \chi_{ac} = 3.1\cdot 10^{-3}$~emu/mol.\\
 \indent The analysis given above can be interpreted as if a fraction of the spins do not reorient in applied magnetic field, but preserve the low-field behaviour described by \eqref{eq:measuredtorque}. This allows us to attempt to obtain better agreement between the experiment and the simulation by assuming reorientation of only a fraction of the spins. We start by dividing the susceptibility tensor in two parts,
\begin{equation}\label{eq:twochi}
	\bm{\hat{\chi}} =  \bm{\hat{\chi}}_1 +\bm{\hat{\chi}}_2.
\end{equation}
In principle, one can expect different tensors $\bm{\hat{\chi}}_1$ and $\bm{\hat{\chi}}_2$ for two subsystems. However, in experiment we can only measure the macroscopic total tensor $\bm{\hat{\chi}}$. So, we try to simulate our data by making a simple reasonable assumption: both subsystems participate in the long range AFM order, and thus their tensors are described by the measured tensor \eref{eq:tensorcsodia}. However, following the result of Ref. \cite{Cvitanic-2018} which claims the magnetic moments on different Cu sites are different, we allow different weights for $\bm{\hat{\chi}}_1 $ and $\bm{\hat{\chi}}_2$. Consequently, we write for tensors of separate subsystems
\begin{align}\nonumber
	\bm{\hat{\chi}}_1&= n \: \bm{\hat{\chi}}_0\: ,\\\label{eq:chi1chi2}
	\bm{\hat{\chi}}_2&= (1-n) \:\bm{\hat{\chi}}_0\:,
\end{align}
where $\bm{\hat{\chi}}_0$ is given by \eqref{eq:tensorcsodia}. The expression \eref{eq:chi1chi2} is written under assumption that both tensors share magnetic eigenaxes. The parameter $n$ describes a contribution of tensor $\bm{\hat{\chi}}_1$ to the total tensor $\bm{\hat{\chi}}_0$, i.e. a contribution of induced magnetization of subsystem 1 to the total magnetization $\mathbf{M} = \mathbf{M}_1 + \mathbf{M}_2 = (	\bm{\hat{\chi}}_1+	\bm{\hat{\chi}}_2)\cdot \mathbf{H}$.  In the following we choose $\bm{\hat{\chi}}_1$ as the susceptibility tensor of the spins that do not reorient in finite magnetic field and $\bm{\hat{\chi}}_2$ as the tensor which is allowed to rotate in an applied magnetic field. Rotation of the tensor $\bm{\hat{\chi}}_2$ is described by \eqref{eq:Rotchi} where instead of $\bm{\hat{\chi}}_0$ we put $\bm{\hat{\chi}}_2$ from expression \eref{eq:chi1chi2}.\\ 
\indent To fix the value of $n$ we calculate again the dependence of magnetization on magnetic field and compare it to the measured values in \fref{fig2}. Since we assume that only spins of the subsystem 2 reorient, now $H_{SF}= [2\:(K_1-K_2)/\Delta \chi_2]^{1/2} $, where $\Delta \chi_2$ represents the susceptibility anisotropy only for subsystem 2 [see \eqref{eq:chi1chi2}]. To reproduce $H_{SF}=1.87$~T at $T=2$~K we set $K_2-K_1=-4.95\cdot10^5$~erg/mol and $n=0.22$. The result of simulation for these parameters is shown by red solid line (simulation 2a) in \fref{fig4}. The spin-flop transition is sharp, as expected for perfect orientation of the sample. Furthermore, our calculation now reproduces measured magnetization values in all applied fields. We expect some misorientation in the experiment (less than 10\degree). Taking this into account we obtain the result shown by the dashed red curve (simulation 2b) in \fref{fig4}. The agreement between the experimental and the calculated data is excellent.\\
\indent We now proceed to simulate the angular dependence of torque measured at $T=4.2$~K assuming reorientation of only fraction of the spins. To reproduce $H_{SF}=1.87$~T we set $K_2-K_1 = - 4.23\cdot 10^{5} $~erg/mol. We also set $n=0.22$ as obtained above and proceed our calculation with the susceptibility tensor values measured at $T=4.2$~K \cite{Zivkovic-2012,Herak-2014}. Furthermore, now it is possible to pinpoint the values of $K_1$ and $K_2$ to obtain the best agreement with the measured torque curves. The above mentioned constraints on anisotropy constants leave two choices for the sign of $K_1$: $K_1>0$ results in an easy axis kind of anisotropy, while $K_1<0$ results in an anisotropy shape that is closer to the easy plane kind of anisotropy. Our data could be well described only by assuming $K_1<0$. From our simulations we find that measured torque for all orientations is best described by the following values of anisotropy constants: $K_1=-0.6\cdot 10^5$erg/mol and $K_2=-4.83\cdot 10^5$ at 4.2~K. The magnetocrystalline anisotropy energy with those anisotropy constants is shown in \fref{fig3}. The result for the $ac$ plane is shown by solid lines in \fref{fig5}. The agreement with experiment is now quite remarkable.\\
\indent In \fref{fig6} we compare the torque measurements for field rotating in the plane spanned by $b$ and $[\overline{1}\:0\:1]^*$ axes to the simulation in the plane spanned by easy and hard axis. Sharp transitions are expected in case of a perfect orientation (dashed curves) while a simulation of small misorientation (solid lines) gives a better agreement with the experiment. In both \fref{fig5} and \fref{fig6}, bottom right panel, we plot the simulated reorientation of the spin axis corresponding to the torque curves measured in different fields. 
Finally, in \fref{fig7} we see excellent agreement between the result of measurement for the plane spanned by $b$ and $[1\:0\:1]^*$ axes and the one obtained by simulation. However, in this plane there is no reorientation in applied field for obtainable values of magnetic field, so results in this plane are not sensitive to whether we allow all spins to rotate or just a fraction of them.
\subsection{\texorpdfstring{Magnetic properties estimated from Density Functional Theory}{Magnetic properties estimated from Density Functional Theory}}\label{sec:DFT}
\begin{figure}[tb]
	\centering
		\includegraphics[width=\columnwidth]{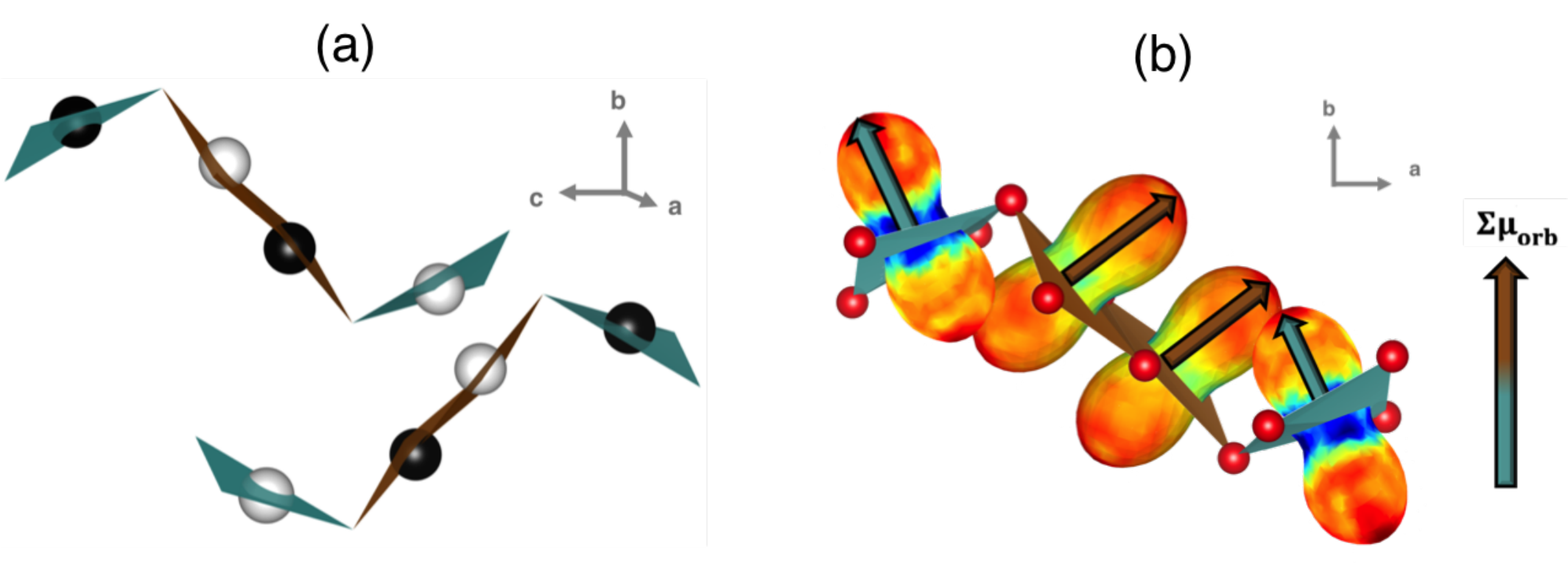}
	\caption{(a) Antiferromagnetic order considered in our DFT calculations. The Cu$^{2+}$ sites are depicted as filled and empty circles, representing up-spin and down-spin, respectively. (b) Orbital moment of one tetramer. Red and blue colors show the maximum and minimum of the orbital moment, respectively. The arrow on the right side shows the vector sum of orbital moments represented by brown and blue vectors on Cu1 and Cu2 sites, respectively.}
	\label{fig8}
\end{figure}

\begin{figure}[tb]
	\centering
		\includegraphics[width=\columnwidth]{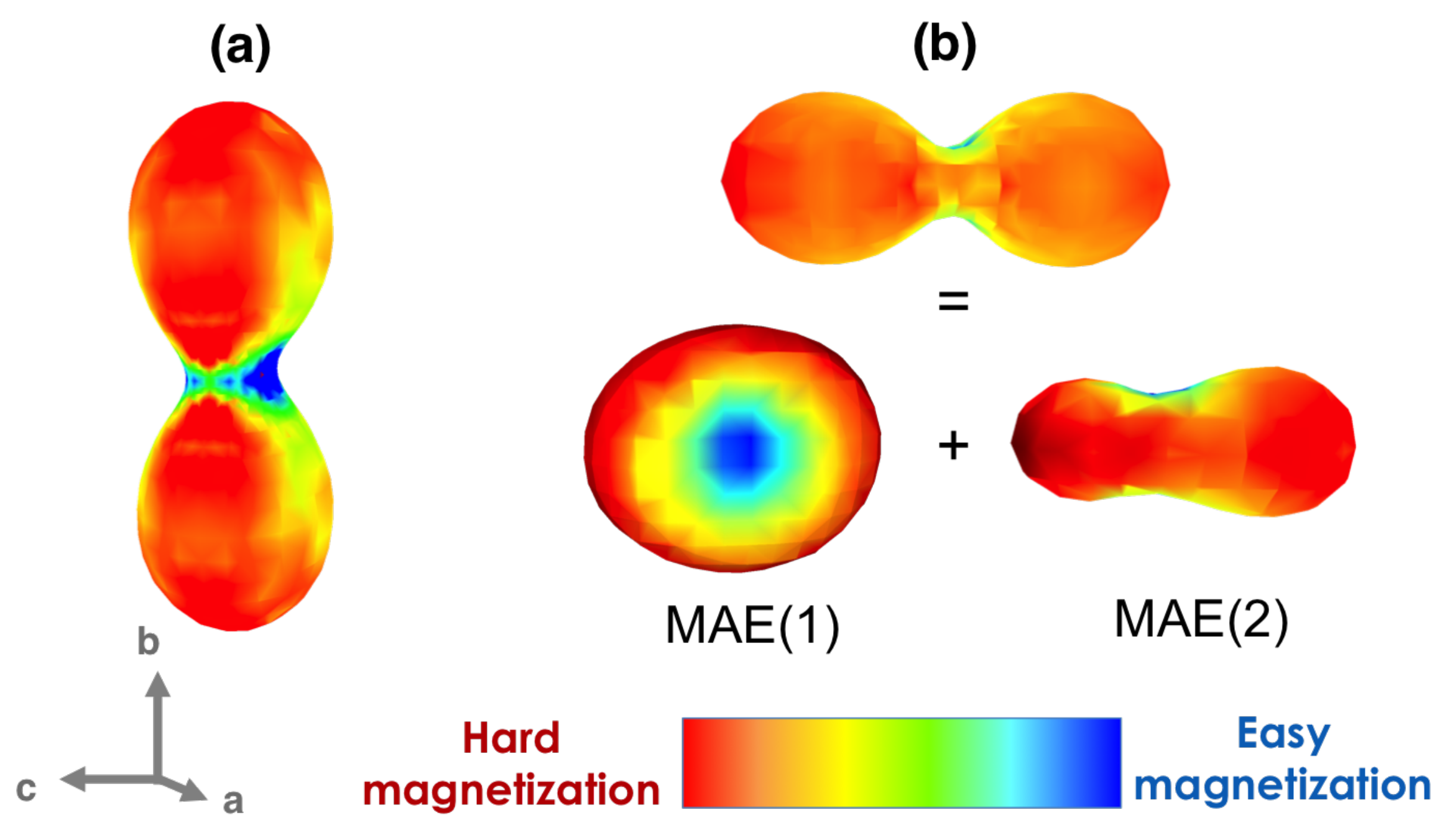}
	\caption{(a) Experimental MAE, (b) Contributions MAE1 (MAE2) to the total MAE (top) determined substituting Cu2 (Cu1) by Zn atoms and considering an $U_{eff}$ = 5 eV correction for Cu sites.}
	\label{fig9}
\end{figure}

\begin{figure}[tb]
	\centering
		\includegraphics[width=\columnwidth]{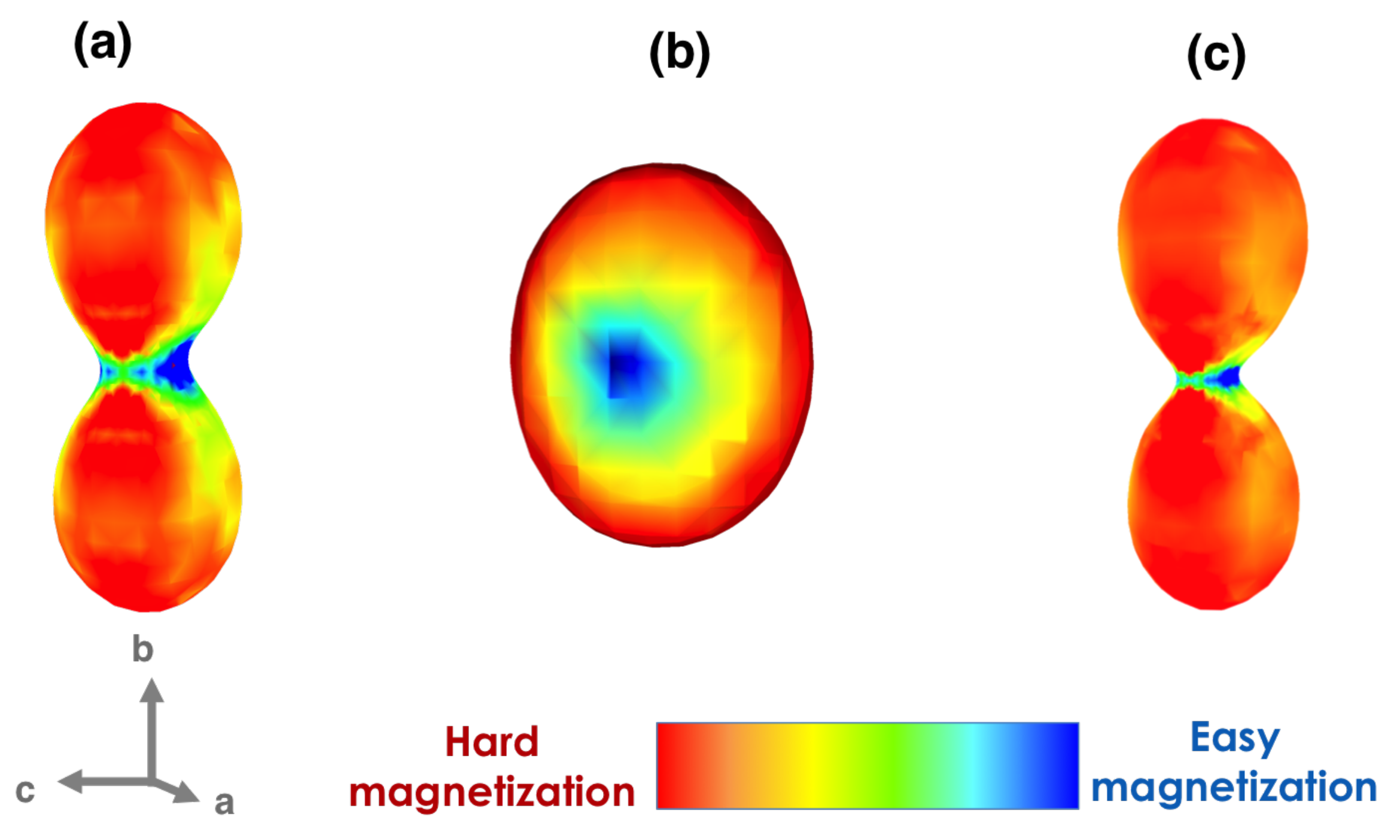}
	\caption{(a) Experimental MAE, (b) MAE considering $U_{eff}$ = 0 eV for Cu1 and $U_{eff}$ = 5 eV for Cu2, (c) similar as previously, except a U - J = 5 - 0.5 eV correction on Cu2.}
	\label{fig10}
\end{figure}

\indent In parallel with our experimental investigations of the magnetic anisotropy of SeCuO$_3$, we have carried out density functional theory (DFT) calculations including spin-orbit coupling (SOC), using a similar strategy to the one we considered for the low-dimensional magnetic compound CuO \cite{rocquefelte_room-temperature_2013}. Indeed, we demonstrated that the estimation of the MAE of CuO, considering its antiferromagnetic ground state, allows to properly predict its easy axis of magnetization. In contrast to CuO, spin fluctuations seem to play a major role in SeCuO$_3$ magnetic properties. \v{Z}ivkovi\'{c} \textit{et al.} have already pointed out the possibility to observe these quantum fluctuations in SeCuO$_3$, and evoked a possible difference of the magnetic moment values between the Cu1 and Cu2 sites \cite{Zivkovic-2012}. More recently, Lee \textit{et al.} mentioned that the site-specific spin correlation may be explained by considering two subsystems based on strongly coupled Cu1 dimers and weakly interacting Cu2 spins \cite{Lee-2017}. They conclude that such a scheme will lead to smaller ordered magnetic moments for Cu1 than for Cu2 due to singlet fluctuations. Also, the reduced ordered magnetic moment of Cu1 sites has been confirmed based on the neutron powder diffraction and NQR measurements \cite{Cvitanic-2018}. However, none of the magnetic models proposed so far allows to explain all the experimental measurements. It is thus essential to provide a theoretical basis to clarify the present picture. \\
\indent Our previous investigation, based on magnetic susceptibility measurements, leads to the conclusion that the Cu2-Cu1-Cu1-Cu2 tetramer is based on two antiferromagnetic couplings, namely J$_{11}$ = 225 K and J$_{12}$ = 160 K \cite{Zivkovic-2012}. Thus, we have generated an antiferromagnetic order noted AF$_1$ shown in \sfref{fig8}{(a)} respecting these conditions, which has been used to estimate the MAE of SeCuO$_3$ with the Wien2k code. \\
\indent SOC is included as a perturbation of the antiferromagnetic collinear state, leading to an energy lowering given by 
	\begin{equation}\label{eq:perturbative-soc}
	\Delta E_{SOC}=\frac { { \left| \left< { i }|{ { \widehat { H }  }_{ SOC } }|{ j} \right>  \right|  }^{ 2 } }{ \left|{ \varepsilon}_{ i }-{ \varepsilon }_{ j }\right| }
	\end{equation}
which accounts for an interaction between an occupied state $i$ with an energy ${ \varepsilon}_{ i }$ and an unoccupied state $j$ with an energy ${ \varepsilon}_{ j }$ via the matrix element $\left< i \: |\widehat{ H }_{ SOC } | j \right>$. The resulting MAE is represented in \sfref{fig9}{(b)}, showing an uniaxial anisotropy along the $b$ direction, while hard axis is in the $ac$ plane. A similar result is observed considering the on-site PBE0 hybrid functional. To be more quantitative, Table \ref{Tab:MAE1} gathered the MAE values for the magnetic eigenaxes of the AFM state (below $T=6$~K), deduced from torque measurements and highlighted in Fig. \ref{fig3}. Given values are expressed relatively to the MAE in the $\left[ 0\: 1\: 0 \right]$ crystal direction. \\
\begin{table}[b]
  \centering
  \caption{MAE values ($\mu$eV /f.u.) for the magnetic eigenaxes of the AFM state, obtained with PBE0, GGA+U with $U_{eff}$ = 9 and 5 eV (noted 9 and 5, respectively) or by substituting zinc for copper ( noted Zn). Directions  $\left[ \overline{1}\: 0\: 1 \right]^{\ast}$ and $\left[ 1 \:0\: 1 \right]$ represent easy and intermediate axes obtained from experiment.}
      \begin{tabular*}{\columnwidth}{@{\extracolsep{\fill} }ccc}
	   \hline	\hline
	   $ U_{eff}^{Cu1}/U_{eff}^{Cu2} $ & $\left[ 1 \:0\: 1 \right]$ & $\left[ \overline{1} \: 0\: 1 \right]^{\ast}$ \\
	   \hline
	   9 / 9 & 4 & 5 \\
	   9 / Zn   & -2 & -1 \\
	   Zn / 9   & 5 & 9 \\
	   \hline
	   5 / 5  & 9 &  9 \\
	   5 / Zn   & -4 & -2 \\
	   Zn / 5 & 11 & 19 \\
	   \hline
	   PBE0 / PBE0 & 11  & 14 \\
		\hline \hline
%    \begin{tabular*}{0.6\textwidth}{@{\extracolsep{\fill} }c|ccc|ccc|c}
%	\hline	\hline
%   $ U_{eff}^{Cu1}/U_{eff}^{Cu2} $    &   9 / 9   &   9 / Zn   &   Zn / 9   &   5 / 5  &   5 / Zn   &   Zn / 5 &   PBE0 / PBE0 \\
%    \hline
%    $\left[ 1 \:0\: 1 \right]$ & 4    & -2  & 5 & 9 & -4 & 11 & 11 \\
%    $\left[ \overline{1} \: 0\: 1 \right]^{\ast}$ & 5    & -1 & 9 & 9 & -2 & 19 & 14 \\\hline\hline
    \end{tabular*}
    \label{Tab:MAE1}
\end{table}
\indent First of all, we have tested two different functionals, i.e. GGA+U and on-site PBE0 hybrid. It appears that GGA+U with $U_{eff}$ = 5 eV leads to similar MAE values to the ones obtained with PBE0. The difference between the easy axis (along $b$) and the two others directions is about 10 $\mu$eV/f.u. If we consider a larger correction ($U_{eff}$ = 9 eV) as in Ref.~\onlinecite{radtke_magnetic_2015} for SrCu$_2$(BO$_3$)$_2$, the MAE values are reduced by a factor of two, but the trend is conserved, i.e. $b$ is still predicted to be the easy axis in disagreement with the experimental facts. To understand this discrepancy, we first consider the Bruno relation \cite{bruno_tight-binding_1989}. According to this model, which is based on the SOC perturbation expression of \eqref{eq:perturbative-soc} and ignoring spin-flip terms, the MAE is directly proportional to the orbital moment anisotropy 
\begin{equation}\label{eq:Bruno}
	MAE = { E }_{ hard }-{ E }_{ easy }=\frac { \xi  }{ 4 } \left| { \left< { L }_{ z } \right>  }_{ hard }-{ \left< { L }_{ z } \right>  }_{ easy } \right| 
\end{equation}
where ${ \left< { L }_{ z } \right>  }$ is the orbital angular momentum. The ${ \left< { L }_{ z } \right> }$ term in \eqref{eq:Bruno} is shown in \sfref{fig8}{(b)} for the four copper sites of one tetramer. For clarity, arrows highlight the direction for which the orbital momentum is maximum, pointing along the normal of each CuO$_4$ plaquette. Summing the orbital moment of all copper sites leads to a total orbital moment which is maximal along the $b$ direction, i.e. an easy magnetization axis is along $b$. Thus, both arguments based on orbital moments and total energy lead to the same conclusion, i.e. an easy axis along the $b$ direction, in contradiction to the experimental data. It should be noticed that as expected for a spin-half system, the dipolar contribution is negligible, less than 0.7 $\mu$eV/f.u.\\
\indent In addition, the present DFT calculations cannot reproduce another experimental fact which is the different magnetic moments at Cu1 and Cu2 sites, 0.46 and 0.73 $\mu_B$, respectively, from NPD \cite{Cvitanic-2018}. Indeed, DFT gives similar magnetic moments for Cu1 and Cu2, i.e. 0.84 and 0.75 $\mu_B$ considering $U_{eff}$ = 9 and 5 eV, respectively. While $U_{eff}$ = 5 eV allows to properly describe the magnetic moment of Cu2, it cannot explain reduced value obtained from experiment for the Cu1. Such a feature appears to be related to the fact that, as in CdCu$_2$(BO$_3$)$_2$, Cu1 ions form strongly coupled singlets, which are polarized by the staggered field of Cu2 spins, and Cu1 and Cu2 magnetic subnetworks are decoupled. \\
\indent We have then estimated the contribution of each inequivalent copper site to the MAE. MAE of Cu1 sites (noted MAE1) was calculated by substituting zinc for copper on all Cu2 sites, and MAE of Cu2 sites (noted MAE2) reversely \cite{weingart_noncollinear_2012}. Indeed, Zn$^{2+}$ and Cu$^{2+}$ cations share nearly the same radii, 0.74 and 0.71 \AA, respectively. In addition, Zn$^{2+}$ is non magnetic (d$^{10}$ electronic configuration) allowing suppression of magnetic response of Zn-substituted sites. A representation of these partial MAE is given in \sfref{fig9}{(b)}. In particular, the easy magnetization axis is located in the $ac$ plane and along the $b$ direction for MAE1 and MAE2, respectively. However, MAE2 is larger in amplitude than MAE1, leading to a total contribution to MAE dictated by MAE2, i.e. an easy axis along the $b$. \\
\indent Our calculations demonstrate the predominant role of the Cu2 subnetwork in the magnetic anisotropy of SeCuO$_3$, but remain incomplete because we do not reproduce the magnetic moment reduction of Cu1. As already mentioned above, such a feature is a consequence of the formation of Cu1 dimers which are in a singlet state at low temperature ($T < 200$ K), the spin fluctuations of Cu1 spins which are different from the ones of Cu2, leading to the decoupling of the two subnetworks, and the staggered field of Cu2 subnetwork which polarizes the magnetic moments of Cu1, leading to a strong decrease of its value \cite{Janson-2012}. \\
\indent From our point of view, all these experimental data converge to one model for SeCuO$_3$, consisting of nearly isolated Cu1 dimers immersed in the staggered field of the AFM long range order of the Cu2 subnetwork. One simple approach is to reduce the Hubbard correction on Cu1 site, and indeed this leads to decrease of Cu1 magnetic moments from 0.84 to 0.75 and 0.60 $\mu_B$, with $U_{eff}$ = 9, 5 and 0 eV, respectively. Reducing $U_{eff}$ even more to negative values will lead to entering an attractive electron-electron interaction regime. Such attractive Hubbard model has been previously used as an effective description for systems involving strong electron-phonon coupling \cite{zitko}. Indeed, strong spin-lattice coupling in the low-temperature state of SeCuO$_3$ has been proposed by Lee \textit{et al.}, based on the measurement of the nuclear spin-lattice relaxation rate 1/T$_1$ \cite{Lee-2017}.\\
\indent Here, the idea is to simulate an extreme situation for which the electron-phonon coupling involving the Cu1 dimers would be enough to overcome the electron-electron Coulomb repulsion. It will correspond to the observation of attractive and repulsive regimes on low and high energy scales, respectively \cite{zitko}. Interestingly, calculating the MAE with two sizable different treatments for Cu1 and Cu2 subnetworks  reproduces the experimental observation. More specifically, we have used $U_{eff}$ = 0 eV for Cu1 and $U_{eff}$ = 5 eV for Cu2. This choice allows us to reproduce in an effective manner the magnetic moment of Cu2 and the reduction of magnetic moments of Cu1. Such treatment leads to a 3D shape shown in \fref{fig10}{(b)}, for which the $b$ direction is properly found as being the hard axis and the easy axis lying in the $ac$ plane. To be more quantitative, we compared in Table \ref{Tab:MAE2} the MAE values for the magnetic eigenaxes of the AFM state with respect to the MAE value along the $b$ axis direction. It now appears that, among these three directions, $[0\:1\:0]$ is systematically the hard one, and $[1\:0\:1]$ the easy one. In other words, by considering that Cu1 and Cu2 subnetworks are decoupled and by taking into account the reduction of the magnetic moment of Cu1, we are able to reproduce the experimental hard magnetization axis along the $b$ direction. The easy axis is found to be in the $ac$ plane in agreement with experimental refinements, but still not in the $\left[ \overline{1}01 \right]^{\ast}$ direction determined from experiment (see \fref{fig10}).\\ 
\begin{table}[tb]
  \centering
  \caption{MAE values ($\mu$eV/f.u.) for the magnetic eigenaxes of the AFM state obtained with different GGA+U treatments for copper atoms. We report values obtained for $U_{eff}$ = 5 eV and $U - J = (5 - 0.5) $eV. Results are given with respect to the MAE along the $b$ axis direction.}
    \begin{tabular*} {\columnwidth}{@{\extracolsep{\fill} }c|cccc}
		\hline \hline 
		 & \multicolumn{2}{c}{$U_{eff}$ } & \multicolumn{2}{c}{$U - J $}\\
    Cu1 / Cu2 & 5 / 5 & 0 / 5 & 5-0.5 / 5-0.5 & 0 / 5-0.5 \\
    \hline
    $\left[ 1\: 0\: 1 \right]$ & 9    & -10     & -7     & -26 \\
    $\left[ \overline{1}\: 0\: 1 \right]^{\ast}$ & 9     & -7     & -20    & -32 \\
		\hline \hline
    \end{tabular*}
    \label{Tab:MAE2}
\end{table}
\indent At this stage, it should be mentioned that Bousquet \textit{et al.} \cite{bousquet_j_2010} have demonstrated that defining explicitely the exchange-correction parameter $J$, in LSDA+U treatment, strongly affects the non-collinear magnetic ground state, and more specifically the spin canting and the magnetocrystalline anisotropy shape. This constitutes a really delicate issue because it implies an adjustment of the amplitude of two parameters, $U$and $J$. Our results for $U - J = 5 - 0.5$ eV are summarized in Table \ref{Tab:MAE2}. It should be noticed that a similar trend of values is obtained using $J = 1$ eV, confirming that the more important aspect is to explicitly specify the $J$ value. Interestingly, the experimental MAE eigenaxes are properly described as soon as $J$ is explicitly defined [see \sfref{fig10}{(c)}]. More specifically, when both Cu1 and Cu2 are corrected, MAE values are 0, -7 and -20 $\mu$eV.fu$^{-1}$ for the $\left[ 0\:1\:0 \right]$,  $\left[ 1\:0\:1 \right]$ and $\left[ \overline{1}\:0\:1 \right]^{\ast}$ directions, respectively. When the correction is added only on Cu2, the MAE values are 0, -26 and -32 $\mu$eV.fu$^{-1}$ for the identical respective directions. Such results demonstrate a drastic change of the MAE, mainly on the intermediate eigenaxes. In such a case, both explicit $J $ definition on Cu2 and reduced Hubbard correction on Cu1 are necessary to properly orientate the theoretical easy axis along the experimental one, as represented in \sfref{fig10}{(c)}. In order to verify how the MAE2 behaves with such treatment, we redo a chemical substitution by Zn atoms on Cu1 sites. As expected, the MAE is strongly modified with respect to the one determined using an $U_{eff}$ treatment [\sfref{fig10}{(b)}], i.e. with an easy, intermediate and hard axes in really good agreement with the experimental ones, as can be witnessed from Figs. \sref{fig10}{(a)} and \sref{fig10}{(c)}. It should be noticed that the overall shape of the MAE was unchanged when considering J = 0.5 and 1 eV for U = 5 eV.
%
%
%%%%%%%%%%%%%%%%%%%%%%%%%%%%%%%%%%%%%%%%%%%%%%%%%%%%%%%%%%%%%%%%%%%%%%%%%%%%%%%%%%%%%%%%%%%%%%%%%%%%%%%%%%%%%%%%%%
%
\section{Discussion\label{sec:disc}}
\begin{figure}[tb]
	\centering
		\includegraphics[width=\columnwidth]{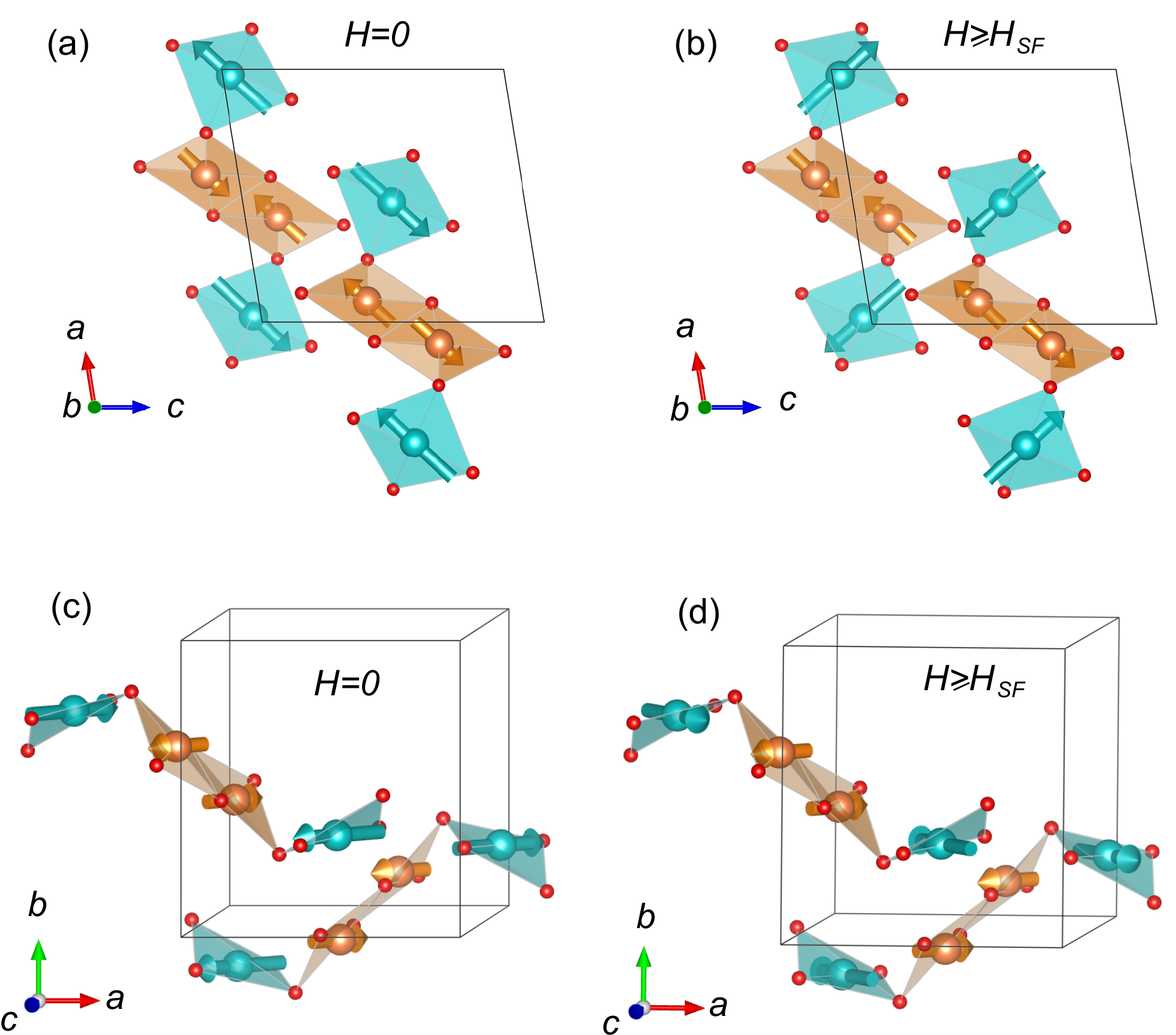}
	\caption{The magnetic structure in AFM state in zero field (a) and (c) and in field $H\geq H_{SF}$ applied along the easy axis (b) and (d) obtained in this work. (a) and (b) show the $ac$ plane to which the spins are confined.}
	\label{fig11}
\end{figure}

\begin{figure}[tbh]
	\centering
		\includegraphics[width=\columnwidth]{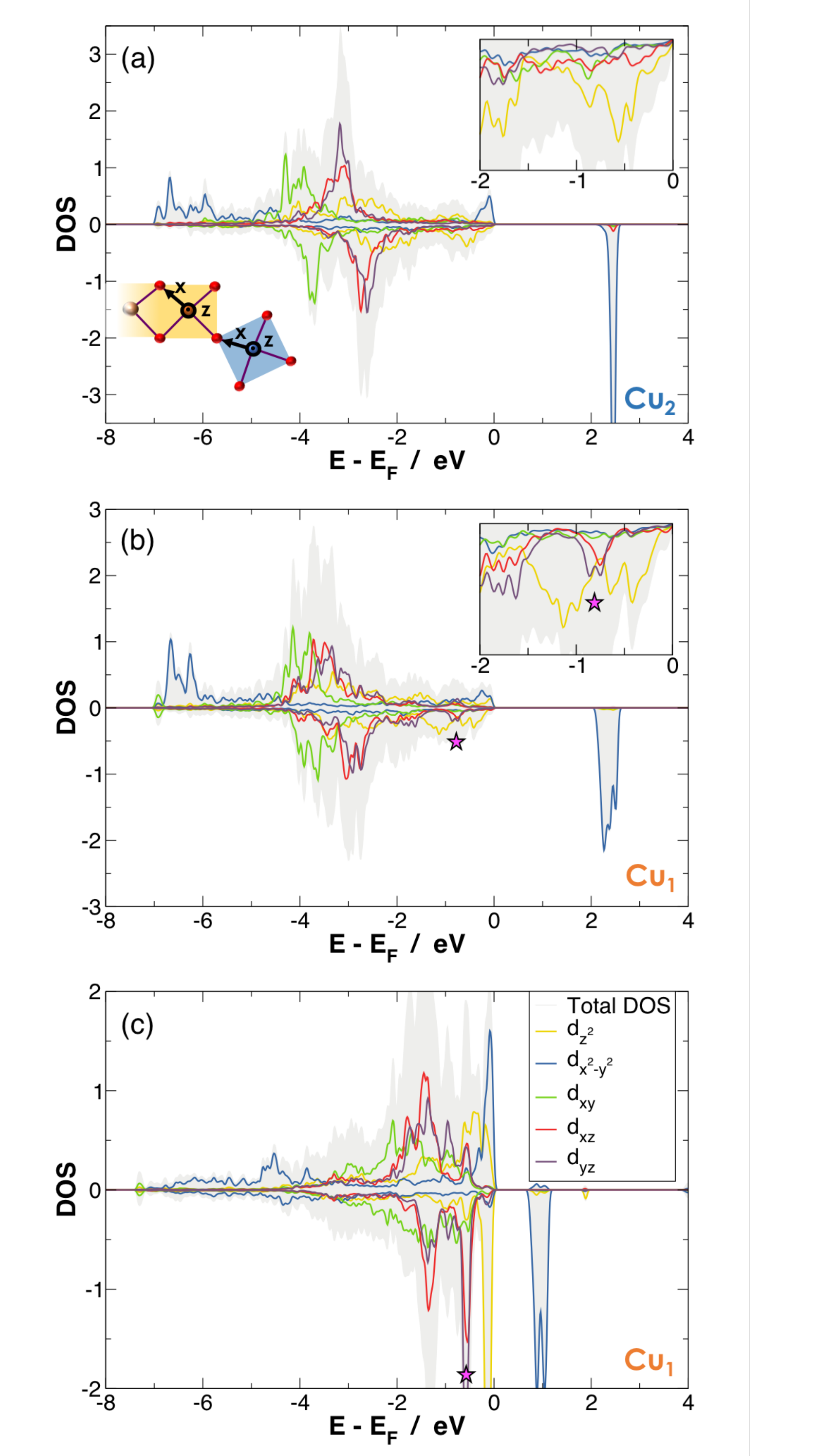}
	\caption{Projected density of states (pDOS) on (a) Cu2 and (b) Cu1 sites considering an $U_{eff}$ = 5 eV, and on (c) Cu1 only at he GGA level. Projection axes for each Cu site is represented on the scheme in (a). Insets zoom the spin minority states over the range of 2 eV below the Fermi energy. Pink stars correspond to first excitations allowed with the excited states.}
	\label{fig12}
\end{figure}

\indent Torque magnetometry is a convenient method for studying magnetocrystalline anisotropy and spin reorientation phenomena in finite magnetic field since angular dependence of torque is very sensitive to the orientation of the spin axis. We employed this method to determine the MAE shape in antiferromagnetic state of SeCuO$_3$. Previous magnetic susceptibility results showed that, below $T_N$, SeCuO$_3$ might be considered a conventional collinear uniaxial antiferromagnet since susceptibility measured along one of the axis goes practically to zero as $T\rightarrow 0$ (easy axis), while along other two axes $\chi$ only slightly increases as the temperature decreases \cite{Zivkovic-2012}. This behaviour is typical for uniaxial collinear antiferromagnets \cite{Neel-1952}, although it does not disqualify a very weak canting of spins. The magnetization measurement at 2~K revealed a spin-flop transition in field of $H_{SF}\approx 1.8$~T applied along the easy axis, also a feature of antiferromagnets with weak magnetocrystalline anisotropy \cite{Neel-1936,Neel-1952}. \\
\indent Using a symmetry allowed MAE and experimentally determined susceptibility tensor we calculated the magnetization as a function of applied field and angular dependence of torque under assumption that all spins collectively rotate in finite magnetic field. This produced results which are expected for uniaxial antiferromagnet, \cite{Neel-1952,Tokumoto-2005,Uozaki-2000}, but which only qualitatively agree with experiment. In $H> H_{SF}$ the magnitude of magnetization should be equal to the value obtained when the field is applied along the intermediate axis, while the experimental value was significantly smaller (see \fref{fig4}). Also, in $H> H_{SF}$ the torque amplitude is expected to become independent on applied magnetic field \cite{Tokumoto-2005,Uozaki-2000}, while in experiment the amplitude showed $H^2$ dependence, but with much smaller slope than in low magnetic field. The $H^2$ dependence of the torque amplitude is characteristic for an AFM system in magnetic field $H\ll H_{SF}$, in which there is no reorientation of spins. This led us to perform simulations where we distinguished two separate subsystems, each represented by its own susceptibility tensor, $\bm{\hat{\chi}}_1$ and $\bm{\hat{\chi}}_2$. The tensor of one subsystem, $\bm{\hat{\chi}}_1$, remains fixed in applied magnetic field, while the tensor of the other subsystem, $\bm{\hat{\chi}}_2$, is allowed to rotate. In this way we obtained quantitative agreement of our simulations with both the magnetization and the torque experiment (see Figs. \ref{fig4} - \ref{fig7}). \\
\indent A possibility of the existence of two subsystems was already mentioned in Ref.~\onlinecite{Zivkovic-2012} where it was suggested that the correlations between Cu1 and Cu2 in SeCuO$_3$ are site-selective and strong coupling between Cu1 spins might form a singlet state at higher temperatures thus separating the Cu1 from the Cu2 spin sublattice. The scenario of two subsystems made of the strongly coupled Cu1 dimers and the weakly coupled Cu2 spins in the AFM state was recently proposed from NMR measurements which witnessed different temperature evolution of $1/T_2$ assigned to Cu1 and Cu2 spins in the AFM state \cite{Lee-2017}. The NQR measurements showed that Cu1 dimers indeed form singlets already at high temperatures $T<200$~K, while Cu2 spins are only weakly coupled to the central pair \cite{Cvitanic-2018}.\\
\indent These results allow us to construct a more rigorous model of the spin reorientation in SeCuO$_3$. Previously published susceptibility anisotropy in the AFM state strongly supports a picture of collinear or very weakly canted AFM state in zero magnetic field \cite{Zivkovic-2012,Herak-2014}. From our macroscopic measurements we are not able to discern the canting so we proceed with the collinear picture. The magnetically ordered state which complies with our torque measurements, both previous \cite{Herak-2014} and from this work, is shown in \sfref{fig11}{(a)} and \sref{fig11}{(c)}. Our theoretical investigations based on a similar magnetic ordering do not allow us to obtain a better picture of this magnetic ordering. Moreover, actual magnetic structure cannot be determined from our macroscopic experiment, only orientation of collinear spins with respect to the crystal axes. To propose a specific orientation of the spins shown in \sfref{fig11}{(a)} and \sref{fig11}{(c)}, we rely on symmetry elements as well as results from literature which allow us to assume AFM coupling between Cu1 spins \cite{Cvitanic-2018} and AFM coupling between Cu1 and Cu2 spins on the tetramer \cite{Zivkovic-2012}. In zero field all spins are oriented along the $\left<\overline{1}\;0\:1\right>^*$ direction (Ref. \cite{Herak-2014} and this work). This result is in disagreement with the neutron powder diffraction data which propose a very noncollinear magnetic structure \cite{Cvitanic-2018}. We point out here that the structure from Ref.~\onlinecite{Cvitanic-2018} is in disagreement with the measured magnetic susceptibility anisotropy \cite{Zivkovic-2012,Herak-2014} which only allows for collinear or very weakly canted spins, as we mentioned above. A large canting proposed in Ref.~\onlinecite{Cvitanic-2018} would not produce a magnetic susceptibility measured along easy axis which goes to zero as $T\rightarrow 0$ \cite{Zivkovic-2012}, a typical feature of collinear or only very slightly canted antiferromagnet.  \\
In order to confirm the present picture, we have realized DFT+U+SOC calculations \cite{liechtenstein_density-functional_1995,perdew_generalized_1996}  using the VASP code \cite{kresse_efficient_1996,kresse_efficiency_1996,kresse_ultrasoft_1999} to take into account the potential non-collinearity. More specifically, we have used an energy cutoff of 550 eV, a similar $k_\textup{mesh}$ to the one in Wien2k and the convergency criterion was fixed at 10$^{-7}$ eV. Starting from a collinear antiferromagnetic arrangement with all the spins oriented along the $\left<\overline{1}\;0\:1\right>^*$, we obtained a small but significant non-collinearity between Cu1 and Cu2 spins, with a canting angle ranging from $0.2$ to $1\degree$ depending on the $U$ (from 5 to 7 eV) and $J$ (from 0 to 1 eV) values. This last result confirms that SeCuO$_3$ can be viewed as a slightly canted antiferromagnet, but the canting is too weak to produce an effect in our macroscopic measurements. It also justifies the Wien2k calculations reported in the present paper, which have been done using a collinear antiferromagnetic model. These calculations have demonstrated that an explicit definition of the exchange-correction term $J$ in the GGA+U+SOC calculations is needed to properly describe the MAE eigenaxes. Moreover, by taking into account, in an effective manner, the different correlation regime of Cu1 and Cu2 subnetworks, we are able to reproduce the reduction of the Cu1 magnetic moment and the relative amplitudes of MAE1 and MAE2.\\
\indent A chemical interpretation of these results can be reached by examining the projected densities of states (pDOS). Indeed, such analysis allows to determine the most important interactions, which are the ones with the smallest energy gap $\left|{ \varepsilon}_{ i }-{ \varepsilon }_{ j }\right|$, as defined in \eqref{eq:perturbative-soc}. More precisely, the observed spin orientations of such Cu$^{2+}$ $S=1/2$ system can be interpreted by the inspection of the pDOS and thus the interactions involving the crystal-field split $d$-states of each magnetic Cu$^{2+}$ ion, under the action of the spin-orbit coupling. Figs.~\sref{fig12}{(a)} and \sref{fig12}{(b)} show the split of Cu1 and Cu2 $d$-states using GGA+U calculations, with $U_{eff}$ = 5 eV. It should be noted that the pDOS obtained with and without specifying explicitly the exchange-correction term $J$ are similar. Here, we defined the local coordinate system with $x $ and $y$ inside the CuO$_4$ plane, pointing towards oxygen atoms, and $z$ perpendicular to the CuO$_4$ plaquette as shown in \sfref{fig12}{(a)}. The overall features for pDOS of Cu1 and Cu2 are the same, with an empty ($x^2-y^2$) state. In such a situation, the $\left< { { d }_{ { xy } } \downarrow}|{ { \widehat { H }  }_{ SOC } }|{ { d }_{ { x }^{ 2 }-{ y }^{ 2 } } \downarrow} \right>$, $\left< { { d }_{ { xz } } \downarrow}|{ { \widehat { H }  }_{ SOC } }|{ { d }_{ { x }^{ 2 }-{ y }^{ 2 } } \downarrow} \right>$ and $\left< { { d }_{ { yz } } \downarrow}|{ { \widehat { H }  }_{ SOC } }|{ { d }_{ { x }^{ 2 }-{ y }^{ 2 } } \downarrow} \right>$ interactions will be non-zero and mainly active because closer to the Fermi level. Spins with orientation inside the plaquette ($\parallel$ $xy$ spin orientation) will be favored if ${ d }_{ { xz } } \downarrow$ or ${ d }_{ { yz } } \downarrow$ states are closer to the empty ${ d }_{ { x }^{ 2 }-{ y }^{ 2 } } \downarrow$ states. In contrast, if ${ d }_{ xy }\downarrow$ states are closer, spins with orientation perpendicular to the CuO$_4$ plaquette ($\perp$ $xy$ spin orientation) will be favored. In the present case, the interpretation is not straightforward due to the significant distortion of Cu1 and Cu2 sites, which are far from regular CuO$_4$ plaquettes. Only Cu1 pDOS show relevant features, which can be interpreted. Indeed, the inset of \sfref{fig12}{(b)} shows that the states which are mainly contributing on the top of the valence band are $d_{xz}$ and $d_{yz}$, which leads to favour the $\parallel$ $xy$ spin orientation, as witnessed in our DFT+U+SOC calculations when considering only Cu1 subnetwork, i.e. MAE1 which shows an easy magnetization axis in the $ac$ plane. In contrast, Cu2 pDOS does allow us to determine which $d$-state is mainly contributing to the top of the valence band. Indeed, while such an analysis is relevant for systems exhibiting regular environments, it should be used with care for distorted environments, because the choice of the local axes for the pDOS is not anymore unique and may influence the results. \sfref{fig12}{(c)} shows the pDOS of Cu1 when considering no Hubbard correction ($U_{eff}$ = 0 eV). The main consequence is a significant band gap reduction and an increase of the $d_{xz}$ and $d_{yz}$ characters on the top of the valence band. Both modifications lead to an increase of the spin-orbit coupling which mixes the $d_{xz}$ and $d_{yz}$ occupied states with the ${ d }_{ { x }^{ 2 }-{ y }^{ 2 } }$ unoccupied state. Such treatment leads to have a larger contribution of MAE1 to the total MAE, which then develops an easy axis in the $ac$ plane and hard axis along the $b$ crystallographic direction. \\
\indent To summarize, there are some similarities between the structure from Ref.~\onlinecite{Cvitanic-2018} and our proposal shown in \sfref{fig11}{(a)} and \sref{fig11}{(c)}. Magnetic moments on Cu2 almost lie within the CuO$_4$ plaquette in both cases and their general direction seems to be similar although it is difficult to make quantitative comparison since authors in Ref.~\onlinecite{Cvitanic-2018} only plotted magnetic structure and did not explicitly provide directions of the magnetic moments. The moments on Cu1 in our structure are neither inside the plaquette nor perpendicular to it, similar to what is proposed by neutron diffraction \cite{Cvitanic-2018}. The direction of moments on Cu1, however, is in our case very different from the one given in Ref.~\onlinecite{Cvitanic-2018}. Our results state that Cu1 moments should be collinear or almost collinear to Cu2 moments, in agreement with the susceptibility, magnetization and torque data, while neutron powder diffraction gives a very noncollinear structure \cite{Cvitanic-2018}. Further investigation of magnetic structure by neutron diffraction on single crystal would resolve this issue.\\
\indent Torque data presented in this work suggest a reorientation for only fraction of the spins in SeCuO$_3$ in magnetic field comparable to the spin-flop field. Present DFT calculations and results from literature \cite{Lee-2017, Cvitanic-2018} allow us to propose that this partial reorientation is in fact site-specific reorientation. Since NMR measurements witness two separate subsystems in AFM state \cite{Lee-2017}, while NQR shows existence of strongly coupled singlet \cite{Cvitanic-2018}, we propose that the susceptibility tensor which rotates in finite magnetic field corresponds to Cu2 spins, while Cu1 spins do not reorient in field applied in our experiment. This results in the magnetic structure shown in \sfref{fig11}{(b)} and \sref{fig11}{(d)} for $H\geq H_{SF}$ applied along easy axis, where spins on Cu2 rotate in the $ac$ plane and are perpendicular to spins on Cu1. This supports the picture where Cu1 and Cu2 subsystems are decoupled in some way. Since half of the spins in SeCuO$_3$ are Cu1 spins, and the other half are Cu2 spins, we can now write for the susceptibility of Cu1 given in \eqref{eq:chi1chi2} $\bm{\chi}_1= n\:\bm{\chi}_0=m\cdot \:(0.5\cdot \bm{\chi}_0)$. In this equation, $0.5\cdot\bm{\chi}_0$ presents the total contribution the Cu1 spins would give to the susceptibility if they were equivalent to the Cu2 spins, and in our model this contribution is reduced by $m$ due to the decoupling of the Cu1 and Cu2 sublattices. For an obtained value $n=0.22$ we get $m=0.44$. The magnetization induced by magnetic field on Cu1 spins is only 44\% of the value it would have if these spins were equivalent to Cu2 spins. It is tempting to compare this to the ratio of magnetic moments $m_{Cu1}/m_{Cu2} \approx 0.4/0.7= 0.57$ obtained for magnetic moments on Cu1 and Cu2 from neutron data \cite{Cvitanic-2018}. However, this comparison may not be justified if Cu1 and Cu2 belong to decoupled subsystems. The ratio $\chi_1/\chi_2=n/(1-n)=0.282$ confirms that the magnetization induced on Cu2 spins is significantly larger than on Cu1 spins. This picture corroborates that the dominant contribution to the total MAE comes from the Cu2 site. \\
\indent The separation of Cu1 and Cu2 subsystems is comparable to the observations in Cu$_2$Cd(BO$_3$)$_2$ \cite{Lee-2014}. The question in SeCuO$_3$ is, are the Cu1 spins polarized singlets in the underlying AFM state formed by interaction between Cu2 spins, or do both Cu1 and Cu2 spins interact mutually to form the AFM state? If Cu1 spins indeed form singlet states even in AFM state, their susceptibility and susceptibility anisotropy should be much smaller than that of the antiferromagnetically ordered Cu2 spins. In fact, if a true singlet state persists in the AFM state, for an intradimer interaction $J\approx 200$~K between Cu1 spins we should expect susceptibility and its anisotropy to be zero at low temperatures \cite{Bleaney-Bowers}. A finite slope of torque amplitude for the non-flopped spins [see \fref{fig2}] suggests Cu1 spins contribute with finite susceptibility anisotropy. Also, $\chi_1/\chi_2=0.282$ obtained from our simulations is small but finite. From our data we cannot distinguish if Cu1 spins in the AFM state form polarized singlets or contribute to the AFM order with a much smaller magnetic moment than Cu2 spins. Site-specific spin reorientation we observe favors the former picture, but further experiments are needed to confirm this.\\
\indent If Cu1 spins are in fact polarized singlets, the neutron diffraction result for Cu2 sublattice \cite{Cvitanic-2018} would then be in agreement with other experimental data as well. For CdCu$_2$(BO$_3$)$_2$ neutron diffraction gave sizable magnetic moment on Cu1 sites \cite{Hase-2009}, while theoretical and experimental data showed that the Cu1-Cu1 dimer in fact forms a singlet in the AFM state \cite{Janson-2012,Lee-2014}. A similar scenario appears to apply to SeCuO$_3$, and indeed our DFT calculations which mimic the reduction of the Cu1 magnetic moment reproduce properly the experimental MAE. Magnetic order in CdCu$_2$(BO$_3$)$_2$ is almost collinear with very small canting, similar to what we propose for SeCuO$_3$. One way to check our proposed magnetic structure in zero and finite field rigorously is to perform single crystal neutron diffraction experiments in zero and applied magnetic field. 
\section{Conclusion}\label{sec:conclusion}
The present paper proposes a combined experimental and theoretical investigation of the magnetic properties of a low-dimensional spin $1/2$ system, which appears to be based on two decoupled magnetic subsystems. This finding, based on measurements on high-quality single crystals and state of the art density functional calculations, opens a way to very exciting physics, with the possibility to control separately two magnetic subsystems in one material. SeCuO$_3$ was previously proposed as a candidate for a system with site-selective spin correlations where Cu1 copper atoms form strongly coupled AFM  dimers, while the coupling including Cu2 spins results in a long range AFM order at low temperatures. Our torque magnetometry results demonstrate site-specific spin reorientation in an applied magnetic field in AFM state of SeCuO$_3$. Using ab-initio approach we show that Cu1 and Cu2 contribute differently to the magnetic anisotropy energy. These results strongly suggest that Cu1 and Cu2 spin systems are decoupled in SeCuO$_3$. Combining our experimental and theoretical findings we propose an antiferromagnetic structure of SeCuO$_3$ in zero field, as well as in field $H \gtrsim H_{SF}$, to be verified by future experiments on this system.
\begin{acknowledgments}
M.~H., N.~N., \v{Z}.~R. and M.~D. acknowledge full support of their work (torque experiment and simulation of torque data) by the Croatian Science Foundation under Grant No. UIP-2014-09-9775. M.~H., M.~D., W.~L.-d.-H. and X.~R.  acknowledge support by COGITO project ''Theoretical and experimental investigations of magnetic and multiferroic materials'' funded by Croatian Ministry of Science and Education and The French Agency for the promotion of higher education, international student services, and international mobility. W.~L.-d.-H. and X.~R. thank the CCIPL (Centre de Calcul Intensif des Pays de la Loire) for computing facilities. The theoretical work was also performed using HPC resources from GENCI- [TGCC/CINES/IDRIS] (Grant 2017-A0010907682). Crystal and magnetic structures in this work were drawn using 3D visualization program VESTA \cite{Vesta}.
\end{acknowledgments}
\appendix*
\section{\label{sec:appendix}Torque magnetometry, susceptibility tensor and magnetic axes}
For a sample with magnetization $\bf{M}$ placed in magnetic field $\bf{H}$, magnetic torque $\bm{\tau} $ acting on the sample can be written as
\begin{equation}\label{eq:torque}
	\bm{\tau} = V\; \mathbf{M} \times \mathbf{H},
\end{equation}
where $V$ is the volume of the sample. For systems in which response of magnetization to magnetic field is linear (such as paramagnets and antiferromagnets in magnetic fields well below the spin-flop field $H_{SF}$) $\bf{M} = \bm{\hat{\chi}} \cdot \bf{H}$, where $\bm{\hat{\chi}}$ is the magnetic susceptibility tensor. Taking into account Neumann's principle and symmetry restrictions for SeCuO$_3$ \cite{Newnham}, the tensor $\bm{\hat{\chi}}$ written in crystal axes coordinate system $(a^*, b, c)$ is given by
	\begin{equation}\label{eq:tensorcso}
	\bm{\hat{\chi}} = 
	\begin{bmatrix}
	\chi_{a^*a^*} & 0 & \chi_{a^*c}\\
	0 & \chi_{bb} & 0 \\
	\chi_{a^*c} & 0 & \chi_{cc}
	\end{bmatrix}.
	\end{equation}
The expression \eref{eq:tensorcso} states that the $b$ axis is one of the eigenaxes of the susceptibility tensor, or magnetic axes, while two other magnetic axes restricted to the $ac$ plane are not necessarily along any of the the crystal axes ($\chi_{a^*c}\neq 0$). \\
\indent The previously published torque magnetometry results showed that in paramagnetic state, apart from the $b$ axis, eigenaxes of the susceptibility tensor in the $ac$ plane are not in the direction of crystal axes \cite{Herak-2014}. Furthermore, torque results showed that magnetic axes in the $ac$ plane rotate as the temperature changes \cite{Zivkovic-2012,Herak-2014}. To understand how this rotation of magnetic axes can be observed in torque measurements, we write the tensor $\bm{\hat{\chi}}$ for the most general orientation of the sample in laboratory coordinate system $(x, y, z)$
\begin{equation}\label{eq:tensorchi}
	\bm{\hat{\chi}} = 
	\begin{bmatrix}
	\chi_{xx} & \chi_{xy} & \chi_{xz}\\
	\chi_{xy} & \chi_{yy} & \chi_{yz} \\
	\chi_{xz} & \chi_{yz} & \chi_{zz}
	\end{bmatrix}.
\end{equation}
In experiment magnetic field is rotated in the $xy$ plane, $\mathbf{H} = H (\cos \varphi \sin \varphi, 0)$, and only the component of torque along the $z$ axis is measured. When tensor $\bm{\chi}$ from \eqref{eq:tensorchi} is introduced in expression \eref{eq:torque} we obtain for the measured component of torque $\tau_z$ 
\begin{equation}\label{eq:torquez}
	\tau_z = \dfrac{m}{2 M_{mol}} H^2 \left[ \left( \chi_{xx} - \chi_{yy}\right)\sin 2\varphi- 2\chi_{xy} \cos 2\varphi \right].
\end{equation}
where $m$ is the mass of the sample given in units of \emph{g}, $M_{mol}$ is the molar mass given in \emph{g/mol} and components of the tensor \eref{eq:tensorchi} are expressed in \emph{emu/mol}. The measured torque is then given in \emph{dyn~cm}. $\varphi$ is the goniometer angle which describes the direction of magnetic field with respect to the laboratory axes. When eigenvectors of the susceptibility tensor are along the $x$ and $y$ axis, the off diagonal component $\chi_{xy}$ is zero and $\chi_{xx} $ and $ \chi_{yy}$ represent the maximal and minimal components of $\bm{\chi}$ in that plane. Otherwise, the maximal and minimal components in the $xy$ plane will be rotated by some angle $\theta_0$ with respect to the $(x, y)$ axis. If we term these new directions $(x', y') $ we can write for the measured torque
\begin{equation}\label{eq:torquemj}
	\tau_z = \dfrac{m}{2 M_{mol}} H^2  \left( \chi_{x'} - \chi_{y'}\right)\sin (2\varphi - 2\varphi_0),
\end{equation}
where correspondence of \eref{eq:torquez} and \eref{eq:torquemj} is established through
\begin{subequations}
\begin{align}
\label{eq:theta0}
\tan(2\varphi_0)&=\dfrac{2\chi_{xy}}{\chi_{xx}-\chi_{yy}},\\
\label{eq:tensorcomp}
\chi_{x'}- \chi_{y'} &= \dfrac{\chi_{xx}-\chi_{yy}}{\cos(2\varphi_0)}=\dfrac{2\chi_{xy}}{\sin(2\varphi_0)}.
\end{align}
\end{subequations}
$\varphi_0$ is the angle that the $x'$ axis makes with the $x$ axis. In experiment the measured component of torque (for paramagnet  and AFM in $H \ll H_{SF}$) can be described by \eqref{eq:torquemj}. From the amplitude of torque we obtain the susceptibility anisotropy in the plane of measurement $xy$ defined as $\Delta \chi_{xy} = \chi_{x'} - \chi_{y'}$. Using equations \eref{eq:theta0} and \eref{eq:tensorcomp} we can determine the difference of the diagonal tensor components $\chi_{xx} - \chi_{yy}$, and the off-diagonal tensor component $\chi_{xy}$.This approach shows that torque magnetometry, in combination with the magnetic susceptibility measurement along one of the directions $x$, $y$ or $z$, allows for determination of the temperature dependence of the susceptibility tensor $\bm{\hat{\chi}}$. By diagonalizing the tensor one can then obtain the temperature dependence of the magnetic susceptibility along magnetic eigenaxes (or magnetic axes for short) and also the orientation of the eigenaxes with respect to the crystal axes. \\
\indent Determining magnetic eigenaxes is crucial in systems where they rotate as temperature changes, as was observed in SeCuO$_3$ in both the paramagnetic and the AFM state \cite{Zivkovic-2012,Herak-2014}, since magnetic eigenaxes also define the MAE shape. Namely, Neumann's principle dictates that the extrema of the MAE are expected to coincide with the magnetic eigenaxes. Our previous results showed that in the AFM state the magnetic axes rotate in the $ac$ plane. Another important result of \eqref{eq:torquez} and \eqref{eq:torquemj} is that the torque measurements can be utilized to observe a symmetry lowering. In case of SeCuO$_3$ this would mean the appearance of finite tensor components $\chi_{a^*b}$ or $\chi_{bc}$ in tensor \eref{eq:tensorcso}. In experiment this means the following. When torque is measured in any plane which contains the $b$ axis, one of the axes $x'$ or $y'$ must correspond to the $b$ axis, i.e. $\chi_{x'}$ or $\chi_{y'}$ must correspond to $\chi_b$, while $\varphi_0$ or $\varphi_0\pm 90\degree$ must correspond to the goniometer angle for which magnetic field is parallel to the $b$. The lowering of symmetry can then be observed as the temperature change of the measured angle $\varphi_0$ in a plane containing the the $b$ axis. For purposes of this work we checked if there is a rotation tha breaks the symmetry in the AFM state in planes containing the $b$ axis. Our results (not shown here) showed that, within experimental error induced by a small misorientation of the sample, there is no symmetry breaking rotation in the AFM state, in agreement with recent NQR and neutron diffraction data \cite{Cvitanic-2018}. At $T=4.2$~K deep in the AFM state, our previous results \cite{Herak-2014} and our new measurements presented here show that the magnetic eigenaxes are ($[\overline{1}\: 0 \:1]^*$, $[1\:0\:1]$, $[0\:1\:0]$).
\bibliographystyle{apsrev4-1}
%\bibliography{Ref}

%
\end{document}